\documentclass[12pt]{article}
\usepackage[top=0.7in,bottom=1in,left=0.8in,right=0.8in,includehead]{geometry}
\usepackage{epsfig,amsmath,amsfonts,verbatim}
\usepackage{hyperref}
\usepackage{lmodern}
\usepackage[T1]{fontenc}
\usepackage{mathtext,marvosym,textcomp}
\usepackage{indentfirst}
\usepackage{epsfig,amsmath,amsfonts}
\usepackage{mathtools}
\usepackage[usenames,dvipsnames,svgnames,table]{xcolor}
\usepackage{tikz}
\usetikzlibrary{arrows}
 \usepackage[ normalem]{ulem}
 \usepackage{tocloft}
  \usepackage{stmaryrd}

\numberwithin{equation}{section}

  \def\g{\gamma} \def\d{\delta} \def\e{\epsilon}
\def\ve{\varepsilon}   
   \def\l{\lambda} \def\m{\mu}
\def\n{\nu} \def\x{\xi}   \def\r{\rho}
 \def\s{\sigma}   \def\f{\varphi}
 \def\c{\chi} \def\y{\psi} \def\w{\omega}

\def\G{\Gamma}

   \def\L{\Lambda} 
 \def\X{\Xi}  
   
\def\F{\Phi}   \def\W{\Omega}

\def\be{\bar{e}}

\def\fr{\frac}  \def\dt{\partial}

\def\mc{\mathcal}

\def\YY{\mathbb{Y}}

\def\TT{\mathbb{T}}

\newcommand\bqa {\begin{eqnarray}}
\newcommand\eqa {\end{eqnarray}}

\newcommand{\bear}{\begin{array}}
\newcommand{\enar}{\end{array}}

\def\beq{\begin{equation}}
\def\eeq{\end{equation}}
\def\bea{\begin{eqnarray}}
\def\eea{\end{eqnarray}}

\def\F{{\mathcal{F}}}

\def\be{\begin{equation}}
\def\ee{\end{equation}}
\def\bfa{{\boldsymbol{\alpha}}}
\def\bfb{{\boldsymbol{\beta}}}

\begin{document}
\renewcommand{\contentsname}{}
\renewcommand{\refname}{\begin{center}References\end{center}}
\renewcommand{\abstractname}{\begin{center}\footnotesize{\bf Abstract}\end{center}} 
 \renewcommand{\cftdot}{}

\begin{titlepage}

\vfill
\begin{flushright}
\end{flushright}

\vspace*{1.6cm}


\begin{center}
   \baselineskip=16pt
   {\large \bf  Fermions and Supersymmetry in E${}_{6(6)}$ Exceptional Field Theory }
   \vskip 2cm
     Edvard T. Musaev$^{* \star}$\footnote{\tt emusaev@hse.ru}, Henning Samtleben$^\star$\footnote{\tt henning.samtleben@ens-lyon.fr}
       \vskip .6cm
             \begin{small}
             			  {\it $^*$ National Research University Higher School of Economics, Faculty of Mathematics \\
             			  7, st. Vavilova, 117312, Moscow, Russia
             			  } \\ [0.5cm]
      					  {\it $^\star$Universit\'e de Lyon, Laboratoire de Physique, UMR 5672, CNRS\\
					  \'Ecole Normale Sup\'erieure de Lyon\\ 46, all\'ee d'Italie, F-69364, Lyon cedex 07, France} \\ [0.5cm]
                        \end{small}
\end{center}

\vskip 1cm

\noindent
\begin{narrower}
We construct the supersymmetric completion of E$_{6(6)}$-covariant exceptional field theory.
The theory is based on a $(5+27)$-dimensional generalized spacetime subject to a covariant section constraint. 
The fermions are tensors under the local Lorentz group ${\rm SO}(1,4)\times {\rm USp}(8)$ and transform as 
weighted scalars under the E$_{6(6)}$ (internal) generalized diffeomorphisms. 
We present the complete Lagrangian and prove its invariance under supersymmetry.
Upon explicit solution of the section constraint the theory embeds full $D=11$ supergravity
and IIB supergravity, respectively.

\end{narrower}

\vfill
\setcounter{footnote}{0}
\end{titlepage}

\clearpage
\setcounter{page}{2}

\tableofcontents

\bigskip
\bigskip
\bigskip

\section{Introduction}

Since the seminal work of Cremmer and Julia \cite{Cremmer:1978ds} it is well known that maximal supergravity compactified on a torus $\TT^d$ enjoys a hidden exceptional symmetry ${\rm E}_{d(d)}$. From the M-theory  point of view these U-duality transformations unify the perturbative T-duality, that relates type IIA and type IIB theories, and the S-duality of type IIB string theory. However, such formulation does not provide a natural geometric interpretation of the duality symmetries.

In the series of works \cite{Hohm:2013pua,Hohm:2013vpa,Hohm:2013uia,Hohm:2014fxa}, exceptional field theory, the ${\rm E}_{d(d)}$-covariant formulation of the full bosonic sector of maximal supergravity, was constructed for $d=6,7,8$. It brings together the ideas from double field theory \cite{Hull:2009mi,Hull:2009zb,Hohm:2010jy,Hohm:2010pp}, its extension to exceptional groups~\cite{Berman:2010is,Berman:2011pe,Berman:2011cg,Berman:2011kg}, and extended geometry~\cite{Coimbra:2011ky,Coimbra:2012af,Aldazabal:2013mya} that is  an extension of Hitchin's generalised geometry~\cite{Hitchin:2004ut, Gualtieri:2003dx} to the case of exceptional duality groups. 
These structures are defined on an exceptional space-time parametrized by external and internal coordinates $\{ x^\mu, \YY^M\}$, $\mu = 0, \dots, 4$; $M=1, \dots, 27$, the latter transforming in the fundamental representation of ${\rm E}_{d(d)}$.\footnote{
In the scheme of~\cite{West:2003fc}, generalized space-time is encoded in the infinite-dimensional $l_1$ representation of ${\rm E}_{11}$.}
 This space is dynamically restricted by a covariant differential constraint called section condition, that allows to systematically drop the extra coordinates and return to the conventional supergravity. The structure of the exceptional field theories resembles the one of the corresponding $(11-d)$-dimensional gauged supergravities~\cite{deWit:2004nw,deWit:2007mt}, however with all fields living on the full exceptional space-time. The dynamics of the ``internal'' sector is formulated in terms of the Lagrangian for a generalised metric, that is constructed from the scalar fields, parametrising the coset space ${\rm G}/{\rm K}$. In this formalism U-duality symmetries are recovered from generalized Lie derivatives in the internal space \cite{Coimbra:2011ky,Berman:2012vc}. Invariance under generalized diffeomorphisms in the external and internal coordinates uniquely fixes all the bosonic couplings of the theory without imposing any supersymmetric structure. Yet, the resulting bosonic system can be supersymmetrized by introducing fermions together with the corresponding connections under the generalized Lorentz group. The supersymmetric version of the full ${\rm E}_{7(7)}$ EFT has been constructed in~\cite{Godazgar:2014nqa}.

In this work we present the supersymmetric completion of the ${\rm E}_{6(6)}$-covariant exceptional field theory that lives on a 5+27-dimensional exceptional space-time. The bosonic theory has been constructed in \cite{Hohm:2013pua,Hohm:2013vpa}. Generalized diffeomorphisms in the internal coordinates $\YY^M$ enter the theory as Yang-Mills type gauge symmetries coupled to the Kaluza-Klein vector field  ${\cal A}_{\mu}{}^M$ in the fundamental representation of $\rm E_{6(6)}$\,.
Fermions enter the theory as spinors under the generalized ${\rm SO}(1,4)\times {\rm USp}(8)$ Lorentz group. Under generalized diffeomorphisms they transform as weighted scalars. As in $D=5$ maximal supergravity \cite{Cremmer:1980gs,deWit:2004nw}, gravitinos $\psi_\mu{}^i$ and fermions $\chi^{ijk}$ transform in the fundamental ${\bf  8}$ and the antisymmetric traceless ${\bf 42}$
representation of ${\rm USp}(8)$, respectively. However, unlike in the five-dimensional truncation, they live on the full exceptional space-time modulo the covariant section condition, which effectively reduces the number of physical coordinates down to ten or eleven. 
Accordingly, the coupling of fermions requires a set of spin connections
\bea
\begin{tabular}{c|ccc}
& $\mc{D}_\m$ & $\mc{D}_M$ \\
\hline \\[-0.4cm]
${\rm SO}(1,4)$ & $\w_\m{}^{ab}$  & $\w_M{}^{ab}$ \\[0.15cm]
${\rm USp}(8)$ & $\mc{Q}_{\m\,i}{}^j$ & $\mc{Q}_{M\,i}{}^j$
\end{tabular}
\qquad
\;,
\label{spinspin}
\eea
in the external and internal directions, and for the two factors of the Lorentz group, respectively.
These connections are defined in terms of the bosonic frame fields, the f\"unfbein $e_\mu{}^a$, and the ${\rm E}_{6(6)}$-valued 27-bein ${\cal V}_M{}^{ij}$.
The ${\rm SO}(1,4)$ connection $\w_\m{}^{ab}$ is defined by the usual vanishing torsion condition 
\bea
\mc{D}_{[\m} e_{\n]}{}^a &=&0 \qquad \Longleftrightarrow\qquad \Gamma_{[\mu\nu]}{}^\rho ~=~0\;,
\eea
however modified by the fact, that the derivative is covariantized also w.r.t.\ internal generalized diffeomorphisms under which the f\"unfbein $e_\mu{}^a$
transforms as a weighted scalar.
For the internal sector on the other hand, vanishing torsion translates into the projection condition~\cite{Coimbra:2011ky}
\bea
\Gamma_{MN}{}^K \Big|_{\bf 351} &=& 0\;,
\eea
for the generalized Christoffel connection, decomposed into irreducible ${\rm E}_{6(6)}$ representations. The off-diagonal blocks in (\ref{spinspin})
finally are determined by demanding that the algebra-valued currents
\bea
{\cal J}_M{}^{ab} &\equiv& e^{a\,\m} {\cal D}[\omega]_M e_\m{}^b\;,\qquad
{\cal J}_{\mu\,kl}{}^{ij} ~\equiv~ 
\mc{V}_{kl}{}^M{\cal D}[{\cal A},{\cal Q}]_\m\mc{V}_M{}^{ij}
\;,
\eea
of the frame fields live in the complement of the maximal compact subalgebra within ${\rm GL}(5)\times {\rm E}_{6(6)}$:
\bea
{\cal J}_M{}^{ab}\Big|_{\mathfrak{so}(1,4)} &=& 0 \;,\qquad
{\cal J}_{\mu\,kl}{}^{ij}\Big|_{\mathfrak{usp}(8)} ~=~ 0 \;.
\eea
Based on these connections we construct the supersymmetry transformation laws and the full supersymmetric Lagrangian in ${\rm E}_{6(6)}$-covariant form.
Upon explicit solutions of the section condition, the Lagrangian reduces to full $D=11$ supergravity and the IIB theory, respectively.

The paper is organized as follows. In section \ref{gauge} we briefly review the structure of the bosonic ${\rm E}_{6(6)}$ exceptional field theory.
We give explicit expressions for the ${\rm SO}(1,4)$ and ${\rm USp}(8)$ connections (\ref{spinspin}) and
the associated curvatures which are the building blocks for the bosonic field equations.
In section~\ref{susy} we present the supersymmetry transformations for all the fields of the theory in a U-duality covariant form
based on the connections~(\ref{spinspin}). The supersymmetry algebra closes with the following schematic form
\begin{equation}
\begin{aligned}
[\d(\e_1),\d(\e_2)]=\;&\x^\m\mc{D}_\m+\d_{\mathfrak{so}(1,4)}(\W^{ab})+\d_{\mathfrak{usp}(8)}(\L^{ij})+\d_{\rm susy}(\e_3)\\
& +\d_{\rm gauge}(\L^M)+\d_{\rm gauge}(\X_{\m\,M})+\d_{\rm gauge}(\X_{\m\n\,\bfa{}})+\d_{\rm gauge}(\Xi_{\m\n\,M})\;,
\end{aligned}
\end{equation}
into the local bosonic symmetries of the theory,
with the explicit transformation parameters listed in \eqref{susy_algebra2} below. The geometry of the extended space deforms the supersymmetry algebra 
in a non-trivial way, although its structural form remains the same as that of the maximal gauged $D=5$ supergravity~\cite{Cremmer:1980gs,deWit:2004nw}.
The full U-duality covariant supersymmetric Lagrangian is then given in section~\ref{invL}.
 In particular, we observe that all Pauli couplings of the fermions to the field strength ${\cal F}_{\mu\nu}{}^M$ can be absorbed by a shift of the internal 
spin connection according to
\bea
\omega^{\pm}_M{}^{ab} &\equiv&
\omega_M{}^{ab} \pm \frac12\,{\cal M}_{MN}\,{\cal F}_{\mu\nu}{}^N\,e^{\mu a} e^{\nu b}\;.
\eea
We sketch the relevant steps in the proof of supersymmetry invariance while the full
calculational details are collected in Appendix~\ref{app:details}.
The results are discussed in section~\ref{sec:conclusions}.

\section{Gauge structure and connections}
\label{gauge}

We start by giving a brief review of the bosonic field content and gauge symmetry 
of the E$_{6(6)}$-covariant exceptional field theory,
constructed in~\cite{Hohm:2013pua,Hohm:2013vpa} (to which we refer for details).
Next we set up the ${\rm USp}(8)\times {\rm E}_{6(6)}$-covariant geometrical formalism
and in particular define the ${\rm SO}(1,4)$ and ${\rm USp}(8)$ spin connections required for
the coupling of fermions. We then work out their various curvatures which are the building blocks for the 
bosonic field equations.

\subsection{Bosonic field content and tensor hierarchy}
\label{tensors}

The bosonic field content of E$_{6(6)}$ exceptional field theory is given by 
\bea
\left\{e_\mu{}^{a}\,,\; {\cal V}_{M}{}^{ij},\; {\cal A}_\mu{}^M\,,\; {\cal B}_{\mu\nu\,M} \right\}
\;,
\label{fieldcontent}
\eea
with indices $\mu, \nu = 0, \dots, 4,$ and $M = 1, \dots, 27,$ labelling external and internal coordinates, respectively,
while indices $a=0, \dots, 4,$ and $i,j = 1, \dots, 8,$ label fundamental indices of the ${\rm SO}(1,4)$ and ${\rm USp}(8)$
Lorentz group, respectively. The f\"unfbein $e_\mu{}^a$ defines the five-dimensional `external' metric as 
$g_{\mu\nu} \equiv e_\mu{}^a e_\nu{}^b \eta_{ab}$ with the flat Minkowski metric $\eta_{ab}$.
Similarly, the pseudo-real 27-bein ${\cal V}_{M}{}^{ij}$ defines an `internal' metric as
\bea
{\cal M}_{MN} &=& {\cal V}_M{}^{ij} {\cal V}_{N\,ij}\;,
\label{defM}
\eea
where ${\cal V}_{M\,ij}\equiv ({\cal V}_M{}^{ij})^*$\,. The 27-bein ${\cal V}_{M}{}^{ij}$ 
can be viewed as an E$_{6(6)}/{\rm USp}(8)$ coset representative 
with the properties
\bea
\mc{V}_M{}^{ij}~=~ \mc{V}_M{}^{[ij]} \;,\qquad \mc{V}_M{}^{ij}\W_{ij}~=~ 0 \;,\qquad
\mc{V}_{M\,ij}~\equiv~(\mc{V}_M{}^{ij})^*~=~\mc{V}_M{}^{kl}\W_{ki}\W_{lj}\;,
\eea
where $\Omega_{ij}=\Omega_{[ij]}$ denotes the symplectic invariant tensor. 
Thus ${\cal M}_{MN}$ in (\ref{defM}) is real and symmetric.
We further define the inverse 27-bein as
\bea
\mc{V}_M{}^{ij}\mc{V}_{ij}{}^N &=& \d_M{}^N\;,\qquad
\mc{V}_M{}^{kl}\mc{V}_{ij}{}^M~=~\d^{kl}_{ij}-\fr18\W_{ij}\W^{kl}
\;,
\eea
where we use conventions $\d^{ij}_{kl}=\frac12(\d^i_k\d^j_l-\d^i_l\d^j_k)$ and $\Omega_{ik}\Omega^{jk}=\delta_i^j$\,.
The fact that the 27-bein is an ${\rm E}_{6(6)}$ group-valued matrix is most efficiently encoded in 
the structure of its infinitesimal variation, 
\bea
\delta {\cal V}_M{}^{ij} &=& 
-2 \,\delta q_k{}^{[i}\,{\cal V}_M{}^{j]k}   + \delta p{}^{ijkl}\,{\cal V}_M{\,}_{kl} 
\;,
\label{varVpq}
\eea
with $\delta q_i{}^{j}$ and $p{}^{ijkl}$ spanning the ${\bf 36}$ and ${\bf 42}$ of ${\rm USp}(8)$,
respectively, i.e.\
\bea
\delta q_i{}^{j}   &=& - \delta q_l{}^{k} \Omega_{ik} \Omega^{jl} \;,\qquad
\delta p{}^{ijkl} ~=~ \delta p{}^{\llbracket ijkl\rrbracket}
\;,
\label{pq}
\eea
and corresponding to the compact and non-compact generators of $\mathfrak{e}_{6(6)}$, respectively.
Double brackets $\llbracket ijkl\rrbracket$ here and in the following 
indicate projection onto the totally antisymmetric and $\Omega$-traceless part, i.e.\
$\delta p^{ijkl}\Omega_{kl}=0$\,.

All fields~(\ref{fieldcontent}) formally depend on the five external coordinates $x^\mu$, and 
27 internal coordinates $\YY^M$, with the latter transforming in the fundamental representation of ${\rm E}_{6(6)}$.
The $\YY^M$-dependence is strongly restricted by the ${\rm E}_{6(6)}$ covariant section 
condition~\cite{Berman:2010is,Berman:2011jh,Berman:2012vc}
\bea
  d^{KMN}\,\partial_M \partial_N A &=& 0\;, \qquad  d^{KMN}\,\partial_M A\, \partial_N B ~=~ 0 \,,
   \label{sectioncondition}
 \eea  
for any fields or gauge parameters $A,B$. 
Here, $d^{KMN}$ is the totally symmetric cubic invariant of ${\rm E}_{6(6)}$.
These constraints admit (at least) two inequivalent solutions, in which the fields
depend on a subset of six or five of the internal coordinates.
The resulting theories are the full $D=11$ supergravity and the type IIB theory, respectively.
For later use we note that
the cubic $E_{6(6)}$ invariant $d^{MNK}$ is related to the symplectic tensor $\W_{ij}$ via
\begin{equation}
\label{d2W}
\begin{aligned}
d^{MNP}=\fr{2}{\sqrt{5}}\mc{V}_{ij}{}^M\mc{V}_{kl}{}^N\mc{V}_{mn}{}^P\W^{jk}\W^{lm}\W^{ni},\\
d_{MNP}=\fr{2}{\sqrt{5}}\mc{V}_M{}^{ij}\mc{V}_N{}^{kl}\mc{V}_P{}^{mn}\W_{jk}\W_{lm}\W_{ni}\;,
\end{aligned}
\end{equation}
as a consequence of the group property of ${\cal V}_{M}{}^{ij}$\,. We use normalization such that
$d_{MKL}d^{NKL}=\delta_M{}^N$\,. With (\ref{d2W}), the section constraint (\ref{sectioncondition})
can be rewritten as
\bea
\Omega_{l[i}{\cal V}_{j]k}{}^M {\cal V}^{kl\,N} \,\partial_M A \partial_N B &=& 
\frac18\,\Omega_{ij}\,{\cal M}^{MN}\, \partial_M A \partial_N B\;,\quad {\rm etc.}
\;,
\label{section2}
\eea
which is a form that we will often use in the following.

The exceptional field theory is invariant under generalized diffeomorphisms in the internal coordinates
which act according to~\cite{Coimbra:2011ky}
\begin{equation}
\label{gen_lie}
(\mc{L}_\Lambda V)^M=\Lambda^N\dt_N V^M-6\,\mathbb{P}^M{}_N{}^K{}_L\dt_K\Lambda ^L V^N+\l_V\dt_P\Lambda^PV^M,
\end{equation}
on a vector $V^M$ of weight $\lambda_V$\,.
Here, $\mathbb{P}^M{}_N{}^K{}_L=(t_{\boldsymbol{\alpha}})_N{}^M (t^{\boldsymbol{\alpha}})_L{}^K$ denotes the projector onto 
the adjoint representation of $E_{6(6)}$,
$(t_{\boldsymbol{\alpha}})_N{}^M$ denoting the representation matrix in the fundamental representation. 
The diffeomorphism parameter $\Lambda^M$ in (\ref{gen_lie}) may depend on internal and external
coordinates. As a result, all external derivatives are covariantized according to
\begin{equation}
D_\m=\dt_\m - \mc{L}_{{\cal A}_\mu}\;,
\label{covD}
\end{equation}
with the vector field ${\cal A}_\mu{}^M$ from (\ref{fieldcontent}).
Accordingly, non-abelian field strengths for vector and two-form fields are defined as
\bea
{\cal F}_{\mu\nu}{}^M &=&  2\, \partial_{[\mu} {\cal A}_{\nu]}{}^M  -2\,{\cal A}_{[\mu}{}^K \partial_K {\cal A}_{\nu]}{}^M 
+10\, d^{MKR}d_{NLR}\,{\cal A}_{[\mu}{}^N\,\partial_K {\cal A}_{\nu]}{}^L
\nonumber\\
&&{}+ 10 \, d^{MNK}\,\partial_K {\cal B}_{\mu\nu\,N}
\;,
\nonumber\\[1ex]
{\cal H}_{\mu\nu\rho\,M} &=&
3\,{ D}_{[\mu} {\cal B}_{\nu\rho]\,M}
-3\,d_{MKL}\, {\cal A}_{[\mu}{}^K\,\partial_{\vphantom{[}\nu} {\cal A}_{\rho]}{}^L 
+ 2\,d_{MKL}\, {\cal A}_{[\mu}{}^K {\cal A}_{\vphantom{[}\nu}{}^P \partial_P {\cal A}_{\rho]}{}^L 
\nonumber\\
&&{}
-10\, d_{MKL}d^{LPR}d_{RNQ}\, {\cal A}_{[\mu}{}^K {\cal A}_{\vphantom{[}\nu}{}^N\,\partial_P {\cal A}_{\rho]}{}^Q
~+~\cdots 
\;,
\label{defFH}
\eea
with the dots 
indicating terms that vanish under projection with $d^{KMN}\partial_N$\,.
Here, vector fields and two-forms carry weight $\lambda_{\cal A}=\frac13$, $\lambda_{\cal B}=\frac23$, respectively,
and the same weight is carried by their respective gauge parameters.
The field strengths (\ref{defFH}) transform covariantly under the non-abelian gauge transformations
 \bea
   \delta {\cal A}_\mu{}^M &=& D_\mu \Lambda^M -10\, d^{MNK} \partial_K\Xi_{\mu N}\;, 
   \nonumber\\
   \delta {\cal B}_{\mu\nu\, M} &=& 2D_{[\mu}\Xi_{\nu]\,M} 
   +d_{MKL}\left(\Lambda^K{\cal F}_{\mu\nu}{}^{L}-{\cal A}_{[\mu}{}^K \delta {\cal A}_{\nu]}{}^L\right)+{\cal O}_{\mu\nu\,M}\;,
 \label{deltaAB}
 \eea
 with $d^{KMN}\partial_M {\cal O}_{\mu\nu\,N}=0$\,. The parameter ${\cal O}_{\mu\nu\,M}$ can be viewed as the tensor gauge
 parameter of the three-forms of the theory which we have not explicitly introduced, since they do not enter the Lagrangian.
 More precisely, this parameter may decomposed according to the field content of three-forms, as
 \bea
 {\cal O}_{\mu\nu\,M} &=& (t^{\boldsymbol{\alpha}})_M{}^N\,\partial_N \Xi_{\mu\nu\,{\boldsymbol{\alpha}}} + \Xi_{\mu\nu\,M}
 \;,
 \label{OXi}
 \eea
 with gauge parameter $\Xi_{\mu\nu\,{\boldsymbol{\alpha}}}$ in the adjoint representation, and a constrained gauge parameter 
 $\Xi_{\mu\nu\,M}$ satisfying the same section condition (\ref{sectioncondition}) as the internal derivatives, i.e.\ 
 $d^{KMN}\,\Xi_{M}\partial_N =0$\,, etc..
This is analogous to the structure of two-forms in ${\rm E}_{7(7)}$ EFT and vector fields in ${\rm E}_{8(8)}$ EFT, respectively, 
c.f.~\cite{Hohm:2013uia,Hohm:2014fxa}.
 The two forms ${\cal B}_{\mu\nu\, M}$ enter the Lagrangian only under projection $d^{KMN}\partial_M {\cal B}_{\mu\nu\,N}$, such that their shift symmetry
 $\delta_{{\cal O}}$ constitutes a trivial symmetry of the action.
 
Under generalized diffeomorphisms, the field strengths ${\cal F}_{\mu\nu}{}^M$ and ${\cal H}_{\mu\nu\rho\,M}$ transform 
 according to (\ref{gen_lie}) as contravariant and covariant vector of weight $\lambda_{{\cal F}}=\frac13$ and 
 $\lambda_{{\cal H}}=\frac23$, respectively.
 In contrast, both are inert under tensor gauge transformations $\Xi_{\mu\,M}$\,.
 The remaining bosonic fields in (\ref{fieldcontent}) transform as scalars under generalized diffeomorphisms with 
 vanishing weight for ${\cal V}_M{}^{ij}$ and weight~$\frac13$ for the f\"unfbein $e_\mu{}^a$\,.

Furthermore, the non-abelian field strengths (\ref{defFH}) satisfy the Bianchi identities
 \bea
  3 \,{ D}_{[\mu}{\cal F}_{\nu\rho]}{}^M &=& 10\, d^{MNK}\partial_K{\cal H}_{\mu\nu\rho\,N}\;, 
  \nonumber\\
4\, { D}_{[\mu} {\cal H}_{\nu\rho\sigma]\,M} &=& 
-3 \,d_{MKL} \,{\cal F}_{[\mu\nu}{}^K {\cal F}_{\rho\sigma]}{}^L
~+~\cdots\;.
\label{Bianchi}
\eea

In addition to the generalized internal diffeomorphisms and tensor gauge transformations (\ref{gen_lie}), 
(\ref{deltaAB}), the theory is invariant under external diffeomorphisms in the coordinates $x^\mu$, 
under which the fields transform as
\bea
 \delta e_{\mu}{}^{a} &=& \xi^{\nu}{ D}_{\nu}e_{\mu}{}^{a}
 + { D}_{\mu}\xi^{\nu} e_{\nu}{}^{a}\;, \nonumber\\
\delta {\cal M}_{MN} &=& \xi^\mu \,{ D}_\mu {\cal M}_{MN}\;,\nonumber\\
\delta {\cal A}_{\mu}{}^M &=& \xi^\nu\,{\cal F}_{\nu\mu}{}^M + {\cal M}^{MN}\,g_{\mu\nu} \,\partial_N \xi^\nu
\;,\nonumber\\
\delta {\cal B}_{\mu\nu\,M} &=& \frac{1}{2\sqrt{10}}\,\xi^\rho\,
 e\varepsilon_{\mu\nu\rho\sigma\tau}\, {\cal F}^{\sigma\tau\,N} {\cal M}_{MN} 
 -d_{MKL}\,{\cal A}_{[\mu}{}^K\, \delta {\cal A}_{\nu]}{}^L
 \;, 
 \label{skewD}
\eea
according to a modified version of the standard five-dimensional diffeomorphisms,
with parameter $\xi^\mu$ which also is a function of $x^\mu$ and $Y^M$\,.

\subsection{Fermions and connections}
\label{subsec:fermions}

The fermionic fields of the theory comprise 8 gravitino fields $\psi_\mu^i$ and 42 spin-$\frac12$ fermions
$\chi^{ijk}=\chi^{\llbracket ijk\rrbracket}$.
With respect to generalized internal diffeomorphisms (\ref{gen_lie}) 
the fermionic fields transform as weighted scalars of weight 
$\lambda_{\psi}=\frac16$, $\lambda_{\chi}=-\frac16$\,.
With respect to the (external and internal) Lorentz group, the fermions are ${\rm SO}(1,4)$ spinors and 
transform in the corresponding representations of ${\rm USp}(8)$.
Like the bosonic fields, also the fermions depend on all coordinates $x^\mu$, $\YY^M$, modulo the section
condition (\ref{sectioncondition}). 
We use the conventions of~\cite{deWit:2004nw} from five-dimensional gauged supergravity.\footnote{
The only exception is our convention for the Levi-Civita density where we follow~\cite{Hohm:2013vpa},
with the two conventions related by
$\varepsilon_{\mu\nu\rho\sigma\tau}^{[1312.0614]}=-i\varepsilon_{\mu\nu\rho\sigma\tau}^{[{\rm hep-th}\slash0412173]}$.
Accordingly, $\gamma$-matrices satisfy $\gamma^{abcde} = i\varepsilon^{abcde}$\,.
}

In particular, we use symplectic Majorana spinors subject to the reality constraint
\begin{equation}
\begin{aligned}
C^{-1}\bar{\y}_i^T&=\W_{ij}\y^j,\qquad
{\y^i}^TC&=\W^{ij}\bar{\y}_j,\qquad
C^{-1}\bar{\c}_{ijk}^T&=\W_{il}\W_{jm}\W_{kn}\c^{lmn},\\
\end{aligned}
\end{equation}
where the charge conjugation matrix $C$ is defined by the following relations
\begin{equation}
\begin{aligned}
C\g_aC^{-1}=\g_a^T, && C^T=-C, && C^\dagger = C^{-1} \;.
\end{aligned}
\end{equation}
This implies the following relation for fermionic bilinears with spinor fields $\y^i$ and $\f^i$ 
\begin{equation}
\bar{\y}_i\G \f^j=-\W_{ik}\W^{jl}\bar{\f}_l(C^{-1}\G^TC)\y^k
\;,
\end{equation}
for any expression of gamma matrices $\G$.

According to the structure of the internal and external Lorentz group there are 
four different blocks of the spin connection 
\begin{equation}
\begin{bmatrix}
\;\w_\m  & {\cal Q}_\m\;\; \\
\;\w_M & {\cal Q}_M\;\;
\end{bmatrix}
\;,
\end{equation}
that ensure ${\rm SO}(1,4)$ and ${\rm USp}(8)$ covariance of external and internal
derivatives, respectively.
Let us discuss them one by one.
The external  ${\rm SO}(1,4)$ connection $\w_\m{}^{ab}$ is defined 
 by the vanishing torsion condition
\begin{equation}
\mc{D}_{[\m} e_{\n]}{}^a \equiv D_{[\m}e_{\n]}{}^a+\w_{[\m}{}^{ab}e_{\n]b}~\stackrel{!}{=}~0\;,
\label{torsion_ext}
\end{equation}
as in standard Riemannian geometry albeit with derivatives $D_\mu$ covariantized according to (\ref{covD}).
Furthermore, the external Christoffel connection $\Gamma_\mu$ can be defined by imposing the vielbein postulate
for the f\"unfbein ${\cal D}_\mu e_\nu{}^a-\Gamma_{\mu\nu}^\lambda e_\lambda{}^a=0$\,.
The internal spin connection on the other hand is defined via
\bea
e^{\mu[a}\,{\cal D}_M e_{\mu}{}^{b]} ~\stackrel{!}{=}~0\quad\Longleftrightarrow\quad
\omega_M{}^{ab} &=& e^{\mu[a} \partial_M e_\mu{}^{b]}
\;.
\label{omega_int}
\eea
Its presence guarantees that internal spinor derivatives transform as 
${\rm SO}(1,4)$ spinors.
As a general notation in the following we will use ${\cal D}$ to indicate (internal or external)
derivatives including all spin connections while $D_\mu$ will only refer to the covariantization (\ref{covD}).
Moreover, in the following it will be useful to define the modified internal spin connections
\bea
\omega^{\pm}_M{}^{ab} &\equiv&
\omega_M{}^{ab} \pm \frac12\,{\cal M}_{MN}\,{\cal F}_{\mu\nu}{}^N\,e^{\mu a} e^{\nu b}
\;,
\label{omegaHat}
\eea
shifted by the non-abelian field strength (\ref{defFH}).
We will denote the corresponding covariant derivatives by ${\cal D}_M^\pm$\,.

Similar relations define the ${\rm USp}(8)$ connections. The external connection ${\cal Q}_{\mu\,i}{}^{j}$
is defined in analogy to $D=5$ gauged supergravity \cite{deWit:2004nw} by imposing that the covariant derivative
of the 27-bein takes the form
\bea
{\cal D}_\mu {\cal V}_M{}^{ij}&\equiv& 
D_\mu {\cal V}_M{}^{ij} +2 \,{\cal Q}_{\mu\,k}{}^{[i}\,{\cal V}_M{}^{j]k}~\stackrel!{=}~
    {\cal P}_\mu{}^{ijkl}\,{\cal V}_M{\,}_{kl} 
\;,
\label{PQ}
\eea
with an ${\rm E}_{6(6)}/{\rm USp}(8)$ coset current 
${\cal P}_\mu{}^{ijkl} = {\cal P}_\mu{}^{\llbracket ijkl \rrbracket}$.
After proper contractions of indices it is straightforward to find the explicit expressions 
\begin{equation}
\begin{aligned}
{\cal Q}_\m{\,}_i{}^j&=\fr13 \mc{V}_{ik}{}^M D_\m \mc{V}_M{}^{jk}\;,\qquad
{\cal P}_\m{}^{ijkl}&= D_\m \mc{V}_M{}^{[ij}\,\mc{V}^{kl]\,M}\;.
\end{aligned}
\label{Q_ext}
\end{equation}
Note the use of the covariant derivative $D_\m=\dt_\m-\mc{L}_{\mc{A}_\m}$ to preserve invariance under 
generalized diffeomorphisms.
These equations imply the following Maurer-Cartan integrability conditions
\begin{equation}
\label{integrability}
\begin{aligned}
{\cal Q}_{\mu\nu\,i}{}^j\equiv
2\,\partial_{[\m}{\cal Q}_{\n]i}{}^j+2\,{\cal Q}_{[\m i}{}^k{\cal Q}_{\n]k}{}^j
&=
-\fr23\,{\cal P}_{[\m iklm}{\cal P}_{\n]}{}^{jklm} -\frac13\,\mc{V}_{ki}{}^M\mc{L}_{{\cal F}_{\m\n}}\mc{V}_M{}^{kj},\\
\mc{D}_{[\m}{\cal P}_{\n]}{}^{ijkl}&=-\fr12\,\mc{L}_{{\cal F}_{\m\n}}\mc{V}_M{}^{[ ij}\mc{V}^{kl] M}\;,
\end{aligned}
\end{equation}
with the field strength ${\cal F}_{\mu\nu}{}^M$ from (\ref{defFH}). It is straightforward to check that the ${\cal B}_{\mu\nu\,M}$ contribution
in the action $\mc{L}_{{\cal F}_{\m\n}}\mc{V}_M{}^{ij}$ drops out due to the section condition~(\ref{sectioncondition}).

Finally, the internal ${\rm USp}(8)$ connection $\mc{Q}_M$ is defined by an analogue of the vanishing torsion condition (\ref{torsion_ext})
for the internal vielbein~\cite{Coimbra:2011ky,Coimbra:2012af}. 
To this end, it is convenient to define the full internal covariant derivative on an ${\rm E}_{6(6)}\times{\rm USp}(8)$ tensor $X_{N}{}^i$ 
of weight $\lambda_X$ as
\bea
\nabla_M X_{N}{}^i &\equiv&
\dt_M X_{N}{}^i - {\cal Q}_{M\,j}{}^i X_N{}^j-\G_{MN}{}^KX_{K}{}^i-\fr34 \l_X \G_{KM}{}^K X_{N}{}^i\;,
\label{nablaM}
\eea
with the algebra valued Christoffel connection $\Gamma_{MN}{}^K \equiv \Gamma_M{}^{\boldsymbol{\alpha}} (t_{\boldsymbol{\alpha}})_N{}^K$\,.
Such defined covariant derivative transforms as a generalized tensor of the weight $\l=\l_X-\fr13$ under generalised diffeomorphisms.
Vanishing torsion corresponds to imposing the relation
\begin{equation}
\mc{T}_{NK}{}^M\equiv\G_{NK}{}^M-6\,\mathbb{P}^M{}_K{}^P{}_L\G_{PN}{}^L+\fr32\,\mathbb{P}^M{}_K{}^Q{}_N\G_{PQ}{}^P ~\stackrel{!}= 0~
\;,
\label{torsion_int}
\end{equation}
which transforms as a tensor under generalized diffeomorphisms (\ref{gen_lie}).
The vanishing torsion condition can equivalently be rewritten as~\cite{Coimbra:2011ky,Aldazabal:2013mya,Cederwall:2013naa}
\begin{equation}
(\mathbb{P}_{351})_M{}^{\bfa N}{}_\bfb \G_N{}^\bfb=0\;,
\label{351}
\end{equation}
with the explicit form of 
the projector $\mathbb{P}_{351}$ onto the $\mathbf{351}$ representation of $E_{6(6)}$ given by
\begin{equation}
(\mathbb{P}_{351})_M{}^{\bfa N}{}_\bfb=-\fr65(t^\bfa)_P{}^N(t_\bfb)_M{}^P+\fr{3}{10}(t^\bfa)_M{}^P(t_\bfb)_P{}^N+\fr15\d^N_M\d^\bfa_\bfb\;.
\end{equation}
A particular consequence of (\ref{351}) is
\bea
d^{MNK}\,\Gamma_{NK}{}^L &=& -\frac12\,d^{MKL}\,\Gamma_{NK}{}^N
\;.
\eea

The vanishing torsion conditions (\ref{torsion_ext}) can be explicitly solved upon imposing the
generalized vielbein postulate for the 27-bein
\bea
\nabla_M \mc{V}_N{}^{ij}&\equiv&
\dt_M  \mc{V}_N{}^{ij} + 2{\cal Q}_{M\,k}{}^{[i} {\cal V}_N{}^{j]k}-\G_{MN}{}^K \mc{V}_K{}^{ij} 
~\stackrel{!}{=} 0
\;,
\label{gvp}
\eea
which allows to express the Christoffel connection in terms of the 27-bein and the internal ${\rm USp}(8)$ connection. 
In turn, the vanishing torsion conditions (\ref{torsion_ext}) translate into the conditions
\bea
\mc{D}_N\mc{V}_K{}^{\llbracket ij } \mc{V}^{kl \rrbracket K} &=&
6\,\mc{D}_K\mc{V}_N{}^{\llbracket ij } \mc{V}^{kl \rrbracket K}-\fr32\,\mc{V}_{N}{}^{\llbracket ij} \mc{V}^{kl\rrbracket M} \,\G_{KM}{}^K\;, 
\label{VT12}\\
\mc{V}_{ik}{}^K\mc{D}_N\mc{V}_K{}^{jk}&=& 3\left(\mc{V}_{ik}{}^M\mc{D}_M\mc{V}_N{}^{jk}- \mc{V}^{jk}{}^M\mc{D}_M\mc{V}_N{}_{ik}\right)
-\fr34\,\G_{KM}{}^K\left(\mc{V}_{ik}{}^M\mc{V}_{N}{}^{jk}-\mc{V}^{jk}{}^M\mc{V}_{N}{}_{ik}\right)\;,
\nonumber
\eea
for the ${\rm USp}(8)$ connection ${\cal Q}_{M\,i}{}^j$\,.
These equations determine (part of) the ${\rm USp}(8)$ connection ${\cal Q}_{M\,i}{}^j$
in terms of the standard decomposition of the Cartan form ${\cal V}^{-1}\partial_M{\cal V}$ 
along the compact and non-compact parts of the 
E$_{6(6)}$ Lie algebra
\bea
\label{Cartan}
q_{M\,i}{}^j &\equiv& \frac13\,{\cal V}_{ik}{}^N \partial_M {\cal V}_{N}{}^{jk}
\;, \qquad
p_M{}^{ijkl} ~\equiv~ 
 \partial_M {\cal V}_N{}^{[ij} {\cal V}^{kl]\,N}
\;.
\eea
Explicitly, parametrizing the connection as
\bea
 {\cal Q}_M{}_j{}^i &=&  {q}_M{}_j{}^i + {\cal V}_M{}^{kl} \Omega^{im} \,q_{kl,jm}
 \;,
 \label{Q_int}
\eea
with $q_{kl,ij}=q_{\llbracket kl\rrbracket ,(ij)}$, it is straightforward to verify that equations (\ref{VT12}) 
are verified provided that\footnote{
An explicit form of ${\cal Q}_M{}_i{}^j$ in terms of the 
${\rm GL}(6)$ components of ${\cal V}_M{}^{ij}$
has been given in~\cite{Coimbra:2011ky}. 
}
\bea
q_{kl,mn} &=&
-p_{M\,klp(m}\, {\cal V}_{n)q}{}^{M} \,\Omega^{pq}
-\frac14\, {\cal V}^{pq\,M} \left(p_{M\,pqk(m}\,\Omega_{n)l}-p_{M\,pql(m}\,\Omega_{n)k} \right)
\nonumber\\
&&{}
+\frac14\,\Gamma_{KM}{}^K\,\left(
{\cal V}_{k(m}{}^M \Omega_{n)l}-{\cal V}_{l(m}{}^M \Omega_{n)k}\right)
+u_{kl,mn}
\;.
\label{q_int}
\eea
Here, $u_{kl,mn}$ denotes the undetermined part of the connection,
satisfying 
\bea
u_{kl,jm}~=~u_{\llbracket kl\rrbracket ,(jm)}\;,\quad
u_{[kl,m]n}&=& 0\;,\quad
u_{kl,jm}\,\Omega^{lj} ~=~0
\;,
\label{conU}
\eea
i.e.\ transforming in the ${\bf 594}$ of ${\rm USp}(8)$, and dropping out from equations (\ref{VT12}).
Vanishing torsion thus determines the ${\rm USp}(8)$ connection (and thereby the Christoffel connection)
up to a block transforming in the ${\bf 594}$ of ${\rm USp}(8)$~\cite{Coimbra:2011ky,Coimbra:2012af,Cederwall:2013naa}.
The undetermined part of this connection drops out of all field equations and supersymmetry variations.
Finally, one may fix the trace part in the Christoffel connection by demanding 
\bea
\nabla_M e &\stackrel{!}{=}&
0\qquad \Longrightarrow\qquad
\Gamma_{KM}{}^K ~=~ \frac45\,e^{-1}\,\partial_M e
\;.
\eea

\subsection{Curvatures}

Let us recollect the notation for the various covariant derivatives
introduced in the previous sections for the external and internal coordinates
\bea
\begin{aligned}
D_\mu &= D[{\cal A}_\nu]_\mu \;,
\\
{\cal D}_\mu &= {\cal D}[A_\nu,\omega_\nu,{\cal Q}_\nu]_\mu \;,&
{\cal D}_M &= {\cal D}[\omega_N,{\cal Q}_N]_M \;,
\\
{\nabla}_\mu &= \nabla[A_\nu,\omega_\nu,{\cal Q}_\nu,\Gamma_\nu]_\mu\;,&\qquad
{\nabla}_M &= \nabla[\omega_N,{\cal Q}_N,\Gamma_N]_\mu\;,
\end{aligned}
\eea
with vector field ${\cal A}_\mu{}^M$ gauging generalized diffeomorphisms as (\ref{covD})
and the composite connections $\omega$, ${\cal Q}$, defined by 
(\ref{torsion_ext}), (\ref{omega_int}), (\ref{Q_ext}), (\ref{Q_int}), in terms of 
the f\"unfbein $e_\mu{}^a$ and the 27-bein ${\cal V}_M{}^{ij}$.
In addition, we recall the modified covariant derivatives ${\cal D}_M^\pm$ and $\nabla_M^\pm$, 
defined with the shifted internal spin connection $\omega_M^\pm$ from (\ref{omegaHat}),
that will come to play their role below.

The external curvature can be evaluated in the standard way by the commutator of covariant derivatives
on an ${\rm SO}(1,4)\times {\rm USp}(8)$ spinor $\epsilon^i$ of weight $\lambda_\epsilon$
\bea
\left[{\cal D}_\mu, {\cal D}_\nu\right] \epsilon^i &=&
\frac14\,{\cal R}_{\mu\nu}{}^{ab}\,\gamma_{ab}\,\epsilon^i
-{\cal Q}_{\mu\nu}{}_j{}^i\,\epsilon^j -{\cal F}_{\mu\nu}{}^M\,\partial_M \epsilon^i
-\lambda_\epsilon\,\partial_M {\cal F}_{\mu\nu}{}^M\, \epsilon^i
\;,
\label{comm_ext1}
\eea
in terms of the Riemann curvature, 
${\rm USp}(8)$ curvature ${\cal Q}_{\mu\nu\,i}{}^{j}$, and the non-abelian
field strength ${\cal F}_{\mu\nu}{}^M$ from (\ref{defFH}).
As it stands however, none of the terms on the r.h.s.\ is simultaneously covariant under generalized
diffeomorphisms and local ${\rm SO}(1,4)\times{\rm USp}(8)$ transformations. In particular, the naive Riemann
curvature defined as the curvature of the external spin connection
\begin{equation}
{{\cal R}_{\mu\nu}}^{ab}=2\,D_{[\mu}\omega_{\nu]}{}^{ab}+2\,\omega_{[\mu}{}^{ac}\,
\omega_{\nu]c}{}^b \;,
\end{equation}
transforms as $\delta_\lambda {\cal R}_{\mu\nu}{}^{ab}= {\cal F}_{\mu\nu}{}^M \partial_M \lambda^{ab}$ under 
${\rm SO}(1,4)$ Lorentz transformations.
Using (\ref{integrability}) and (\ref{gvp}), the terms on the r.h.s.\ of (\ref{comm_ext1}) can be rearranged
into the manifestly covariant expressions
\bea
\left[{\cal D}_\mu, {\cal D}_\nu\right] \epsilon^i &=&
\frac14\,\widehat{\cal R}_{\mu\nu}{}^{ab}\,\gamma_{ab}\,\epsilon^i
+\fr23\,{\cal P}_{[\m jklm}{\cal P}_{\n]}{}^{iklm}\,\epsilon^j
+ \nabla_{M} \F^N  \left(
{\cal V}_N{}^{jk} {\cal V}_{ik}{}^M-
 {\cal V}_{N\,ik} {\cal V}^{jk\,M} \right)\epsilon^j 
\nonumber\\
&&{}
-{\cal F}_{\mu\nu}{}^M\,\nabla_M \epsilon^i
-\lambda_\epsilon\,\nabla_M {\cal F}_{\mu\nu}{}^M\, \epsilon^i
\;,
\label{comm_ext}
\eea
with the full covariant internal derivatives $\nabla_M$ from (\ref{nablaM}) and the
`improved' Riemann tensor defined by~\cite{Hohm:2013jma,Hohm:2013vpa}
\bea
\widehat{\cal R}_{\mu\nu}{}^{ab} &\equiv & {\cal R}_{\mu\nu}{}^{ab}
+ {\cal F}_{\mu\nu}{}^M\,\omega_M{}^{ab}
\;,
\label{RiemannHat}
\eea
transforming covariantly under local Lorentz transformations.
For later use, we note that this tensor and the associated Ricci tensor
$\widehat{\cal R}_{\mu\nu}\equiv {\cal R}_{\mu\rho}{}^{ab}e_a{}^\rho e_{\nu b}$,
satisfy the modified Bianchi identities
\bea
\widehat{\cal R}_{[\mu\nu]} &=& -\frac12\, g_{\rho[\mu} \nabla_M {\cal F}_{\nu]}{}^{\rho\,M}\;,
\nonumber\\
\widehat{{\cal R}}_{[\mu\nu\rho]}{}^{a} 
 &=&  
-    {\cal F}_{[\mu\nu}{}^K \,\nabla_K e_{\rho]}{}^{a} 
 - \frac13\, e_{[\mu}{}^{a}  \nabla_K {\cal F}_{\nu\rho]}{}^K 
 \;.
 \label{BianchiR}
\eea
In contrast, the symmetric part of the Ricci tensor $\widehat{\cal R}_{(\mu\nu)}$ 
will appear in the Einstein field equations
in the standard way.
Similar to (\ref{comm_ext}), the Maurer-Cartan integrability relations for the
coset currents (\ref{integrability})
can be rewritten in the manifestly covariant form
\begin{equation}\begin{aligned}
{\cal D}_{[\mu} {\cal P}_{\nu]}{}^{ijkl} &=& -3\, {\cal V}_N{}^{\llbracket ij}{\cal V}^{kl\rrbracket M} \,\nabla_M\F_{\mu\nu}{}^N
\;.
\label{DPcov}
\end{aligned}\end{equation}

Let us now discuss the mixed components of the curvature, i.e.\ the tensors obtained by commuting
internal with external covariant derivatives. We only consider combinations of commutators
in which the undetermined part of the ${\rm USp}(8)$ spin connection (\ref{Q_int}), (\ref{q_int}) drops out.
This is the case for
\bea
\label{comm_1}
{\cal V}_{ij}{}^M \left[{\cal D}^-_M,{\cal D}_\mu \right] \epsilon^j &=&
\frac12\,{\cal V}^{jk}{}^M {\cal D}_M {\cal P}_{\mu}{\,}_{ijkn} \epsilon^n 
+\frac14\,{\cal R}^-_{M\mu}{}^{ab}\,\gamma_{ab}\,\epsilon^j
\;.
\eea
Indeed, the undetermined connection on the l.h.s.\ appears as 
$\Omega^{jm} u_{ij,km} =0$. On the r.h.s.,
the two term describe the mixed ${\rm USp}(8)$ and ${\rm SO}(1,4)$ curvature,
respectively, with the second term defined by the tensor
\bea
{\cal R}^-_{M\mu}{}^{ab} &\equiv& 
\partial_M \, \omega_\mu{}^{ab}
-{\cal D}_\mu \,\omega^-_M{}^{ab}
\;.
\label{Rmix}
\eea
Evaluating this curvature gives rise to its Bianchi identity
\bea
{\cal R}^-_{M[\nu\,\rho\sigma]}
&=& 
\frac12\, 
{\cal D}_{[\nu} 
\left({\cal F}_{\rho\sigma]}{}^N\,{\cal M}_{NM} \right)
\;,
\label{BianchiMix}
\eea
and the mixed Ricci tensor
\bea
{{\cal R}}^-_{M\nu}{}^{\mu\nu} &=&
-\frac12\, \widehat{J}^\mu{}_M +\frac12\,e_{a}{}^\mu e_{b}{}^\nu\,
 {\cal D}_\nu \left( {\cal M}_{MN} {\cal F}^{ab}{}^N \right)
\;,
\label{BIR2}
\eea
with the current $\widehat{J}^\mu{}_M$ defined by
\bea\label{STep}
\widehat{J}^\mu{}_{M} &\equiv&
-2  e_{a}{}^\mu e_{b}{}^\nu \left(
\partial_M \omega_{\nu}{}^{ab} 
-  {\cal D}_{\nu} \left(   e^{\rho[a} \partial_M e_{\rho}{}^{b]} \right) \right)
\;,
\eea
that will feature in the vector field equations.
Similar to (\ref{comm_1}), we can evaluate the following combination of commutators
\bea
 \mc{V}^{\llbracket ij \, M}\left[{\cal D}^-_M,{\cal D}_\mu \right] \epsilon^{k\rrbracket}&=& 
-2\,{\cal V}_{mn}{}^M\,\Omega_{pr} \Omega^{m\llbracket i} {\cal D}_M{\cal P}_{\mu}{}^{jk\rrbracket np} \epsilon^{r}
-\frac12\,{\cal V}_{mn}{}^M\,{\cal D}_M{\cal P}_{\mu}{}^{mn\llbracket ij} \epsilon^{k\rrbracket}
\nonumber\\
&&{}
+\frac14\,{\cal R}^-_{M\mu}{}^{ab}  \mc{V}^{\llbracket ij\,M}\,\gamma_{ab}\e^{k\rrbracket}
\;,
\label{comm2}
\eea
in terms of the coset current ${\cal P}_\mu{}^{ijkl}$ and the mixed curvature (\ref{Rmix}).
Again, the undetermined part of the ${\rm USp}(8)$ spin connection drops out on the l.h.s..

Let us finally discuss the internal components of the curvature. These are obtained from commutators
of internal derivatives in combinations such that the undetermined part of the spin connections drops out.
The relevant combinations are given by~\cite{Coimbra:2011ky,Coimbra:2012af,Aldazabal:2013mya}
\bea
{\cal V}^{ik\,M} {\cal V}_{kj}{}^{N}\left[\nabla_M, \nabla_N\right] \e^j+
\left(4{\cal V}^{ik\,M} {\cal V}_{kj}{}^{N} + \frac12{\cal M}^{MN}\,\delta_j^i \right)
\nabla_{(M} \nabla_{N)} \e^j
&=&
\frac14\,{\cal V}^{ik\,M} {\cal V}_{kj}{}^{N}\,{\cal R}_{MN}{}^{ab}\,\gamma_{ab}\,\e^j
\nonumber\\
&&{} 
-\frac1{16}\,{\cal R}\,\e^i\;,
\label{comm_nabla_1}\\[2ex]
{\cal V}^{\llbracket i j\,N} {\cal V}^{k\rrbracket l\,M}  \Omega_{ln} [\nabla_M, \nabla_N]\,
 \e^n
 +2\, \Omega_{l m} {\cal V}^{l \llbracket i \,M} {\cal V}^{j |m|\,N}  
 \nabla_{(M} \nabla_{N)}\, \e^{k\rrbracket} 
&=&
\frac14\,{\cal V}^{\llbracket i j\,N} {\cal V}^{k\rrbracket l\,M} \Omega_{ln} {\cal R}_{MN}{}^{ab}\gamma_{ab}\e^n
\nonumber\\
&&{}-\frac14\, {\cal R}^{ijkl} \, \Omega_{ln} \,\e^n
\;.
\label{comm_nabla_2}
\eea
The combinations on the l.h.s.\ are such that the undetermined part $u_{kl,mn}$ of the ${\rm USp}(8)$ spin connection is projected out
while the leading two-derivative terms vanish due to the section condition (\ref{section2}).
The first terms on the r.h.s.\ refer to the curvature of the internal spin connection (\ref{omega_int})
which takes the form~\cite{Godazgar:2014nqa}
\bea 
 {\cal R}_{MN}{}^{ab}
&=& 
  - \frac{1}{2} \, e^{\mu[a}e^{b]\nu}g^{\sigma \tau} \nabla_{M} g_{\mu \sigma} \nabla_{N} g_{\nu \tau} 
  \;.
  \label{RMNab}
\eea
The generalized scalar curvatures ${\cal R}$ and ${\cal R}^{ijkl}$ in (\ref{comm_nabla_2}) can be evaluated using the explicit 
expressions for the ${\rm USp}(8)$ spin connection (\ref{Q_int}), (\ref{q_int}), leading to
\bea
{\cal R} &=&
-2\,{\cal V}_{ij}{}^M {\cal V}_{kl}{}^N \,\left(\partial_M p_{N}{}^{ijkl}
+ 4q_M{}_m{}^{[i} p_N{}^{jkl]m}\right)
+ \frac16\, {\cal M}^{MN}\,p_{M\,ijkl} p_{N}{}^{ijkl}
\nonumber\\
&&{}
+ 2\,{\cal V}_{ij}{}^M {\cal V}^{kl\,N}\,p_{M}{}^{ijmn} \,p_{N\,klmn}
- \frac{16}{5}\,{\cal V}_{ij}{}^M {\cal V}_{kl}{}^N \,e^{-1}\,\partial_M e\,p_N{}^{ijkl}
\nonumber\\
&&{}
+\frac{8}{5}\,{\cal M}^{MN} e^{-1} \partial_M \partial_N e 
-\frac{4}{5}\,{\cal M}^{MN}\,e^{-2}\,\partial_M e \partial_N e 
\;,\nonumber\\[2ex]
{\cal R}^{ijkl}&=&  
 \frac{1}{3}\,  {\cal M}^{MN}\,e^{-1}\left(
 {\partial}_{M}(e {p}_{N}\,^{i j k l}) + 4\,e q_{M\,m}{}^{\llbracket i} p_N{}^{jkl\rrbracket m}\right)
\nonumber\\
&&{}
- 4\,  \mc{V}_{mn}{}^M  e^{-1}\left(
{\partial}_{M}(e {p}_{N}\,^{m n\llbracket ij} )
+2\,e  p_N{}^{mn p\llbracket i} q_{M\,p}{}^{j}
+2\,e q_{M\,p}{}^{m} p_N{}^{np\llbracket ij}\right) {\cal V}^{kl \rrbracket N}
\nonumber\\
&&{}
+ 2\,  {\cal V}^{\llbracket ij M} {\cal V}^{kl\rrbracket N} e^{-1} {\partial}_{M} \partial_N e 
- \frac{8}{5}\,  {\cal V}^{\llbracket ij M} {\cal V}^{kl\rrbracket N}  {e}^{-2}\,{\partial}_{M}e {\partial}_{N}e  
\nonumber\\
&&{}
- \frac{2}{3}\,  \mc{V}_{mn}{}^M \mc{V}_{pq}{}^N  {p}_{M}{}^{i j k l} {p}_{N}{}^{m n p q} 
+ \frac{32}{3}\,  {\cal V}_{mn}{}^{[M} {\cal V}_{pq}{}^{N]}  {p}_{M}{}^{m n p\llbracket i} {p}_{N}{}^{j k l\rrbracket q} 
\nonumber\\
&&{}
+4\,  \mc{V}_{mn}{}^M \mc{V}_{pq}{}^N  {p}_{M}{}^{m n\llbracket ij} {p}_{N}{}^{k l\rrbracket p q} 
+ \frac{1}{3}\,  {\cal V}^{\llbracket ij M} {\cal V}^{kl\rrbracket N}  {p}_{M\, m n p q} \,{p}_{N}{}^{m n p q}
\;,
\label{Rijkl}
\eea
in terms of the 27-bein and its derivatives. 
Their explicit calculation requires a number of non-trivial ${\rm USp}(8)$ identities, some
of which are collected in appendix~\ref{app:identities}.
Together with (\ref{RMNab}), these curvatures appear in the
Einstein and the scalar field equations, respectively.
For the following, it is also useful to note the relation between the curvature components
${\cal R}$ and ${\cal R}^{ijkl}$: under a non-compact $\mathfrak{e}_{6(6)}$ transformation
of the form
\begin{equation}\begin{aligned}
\delta {\cal V}_{M\,ij} &=& - \Sigma_{ijkl}\,{\cal V}_{M}{}^{kl}\;,
\end{aligned}\end{equation}
the scalar curvature ${\cal R}$ transforms as
\begin{equation}\begin{aligned}
\delta {\cal R} &=&\,{\cal R}^{ijkl}\,\Sigma_{ijkl} ~+~\nabla_M {\cal J}_\Sigma^M 
\;,
\label{varR1}
\end{aligned}\end{equation}
into the ${\cal R}^{ijkl}$ curvature, up to a boundary current of weight $\lambda_{{\cal J}_\Sigma}=-\frac13$\,.
Moreover, the dependence of ${\cal R}$ on the external metric is such that
\bea
\delta (e{\cal R}) &=& (\delta e) \,{\cal R}~+\mbox{total derivatives}\;.
\label{varR2}
\eea

\section{Supersymmetry transformations and algebra}
\label{susy}

As the main result of this section we present the supersymmetry transformation rules for all the fields of 
the $E_{6(6)}$ exceptional field theory and verify that their algebra consistently closes into generalized diffeomorphisms and
gauge transformations. The full set of supersymmetry transformations is given by
\begin{equation}
\label{transf}
\begin{aligned}
\d_\e \y_\m^i&=\mc{D}_\m \e^i-i\sqrt{2}\,\mc{V}^{ij\,M}\left({\nabla}^-_M(\g_\m\e^k)-\fr13\,\g_\m{{\nabla}}^-_M\e^k\right)\W_{jk}\;,\\
\d_\e\c^{ijk}&=\fr i2\, {\cal P}_\m{}^{ijkl}\W_{lm}\, \g^\m\e^m
+ \fr{3}{\sqrt{2}} \,\mc{V}^{\llbracket ij\,M} \,{\nabla}^-_M\e^{k\rrbracket}\;,\\[2ex]
\d_\e e^a_\m&=\fr12\,\bar{\e}_i\g^a\y_\m^i\;,\qquad
\d_\e\mc{V}_M{}^{ij} =4i\,\W^{im}\W^{jn}\,\mc{V}_M{}^{kl}\,\W_{p\llbracket k}\bar{\c}_{lmn\rrbracket}\e^p\,,\\
\d_\e {\cal A}_\m{}^M&=\sqrt{2}\left(i\,\W^{ik}\bar{\e}_k\y_\m{}^j+\bar{\e}_k\g_{\m}\c^{ijk}\right)\mc{V}_{ij}{}^M,\\
\d_\e {\cal B}_{\m\n}{}_M&=-\fr{1}{\sqrt{5}}\,\mc{V}_{M}{}^{ij}\left(2\,\bar{\y}_{i[\m}\g_{\n]}\e^k\W_{jk}+i\bar{\c}_{ijk}\g_{\m\n}\e^k\right)
- d_{MNP}\,{\cal A}_{[\m}{}^N\d_\e  {\cal A}_{\n]}{}^P\;,
\end{aligned}
\end{equation}
in terms of the covariant derivatives defined above. Spinor conventions were summarized in section~\ref{subsec:fermions}.
Upon dropping all internal derivatives $\partial_M\longrightarrow0$, these transformation rules precisely reproduce those
of $D=5$ maximal supergravity~\cite{Cremmer:1980gs,deWit:2004nw}.\footnote{
To be precise, we note the rescaling of gauge and tensor fields ${\cal A}_\m{}^M{}_{[1312.0614]}= \frac1{\sqrt{2}}{A}_\m{}^M{}_{[{\rm hep-th}\slash0412173]}$, 
${\cal B}_{\m\n\, M}{}_{[1312.0614]}=-\frac14{B}_{\m\n\, M}{}_{[{\rm hep-th}\slash0412173]}$ together with rescaling of the associated symmetry parameters,
in order to translate the notation from~\cite{deWit:2004nw} into~\cite{Hohm:2013vpa}.
In this paper, we will stick to the conventions of~\cite{Hohm:2013vpa} for the normalization of the gauge fields.}
It is interesting to note that just as for the supersymmetric ${\rm E}_{7(7)}$ theory~\cite{Godazgar:2014nqa}, all appearance of the 
gauge field strength $\mc{F}_{\m\n}{}^M$ in the transformation rules can be absorbed into the homogeneous shift (\ref{omegaHat}) of the
internal spin connection.
In the next section, we will see that the supersymmetric Lagrangian in contrast carries the opposite derivative $\nabla_M^+$ as well.

The internal derivatives $\nabla_M$ appear in the supersymmetry transformations
only in particular combinations such that the undetermined part of the ${\rm USp}(8)$ connection (\ref{Q_int}) drops out~\cite{Coimbra:2011ky,Coimbra:2012af}.
With the explicit parametrization of ${\cal Q}_{M\,i}{}^j$ from (\ref{q_int}) we may explicitly evaluate these derivatives in terms of the Cartan form (\ref{Cartan})
of the 27-bein as
\bea
{\cal V}_{ij}{}^M {\nabla}_M \epsilon^j &=& \mc{V}_{ij}{}^M\left(\partial_M\e^j-q_{M\,k}{}^j \e^k\right)
-\frac12\,{\cal V}^{jk}{}^M p_{M}{}_{ijkn} \epsilon^n 
+\frac14\,(2-3\lambda_\epsilon)\, {\cal V}_{ij}{}^M \Gamma_{KM}{}^K \,\epsilon^j
\;,
\nonumber\\
\mc{V}^{\llbracket ij\,M}{{\nabla}}_M\e^{k\rrbracket} &=& 
\mc{V}^{\llbracket ij\,M}\left( {{\partial}}_M\e^{k\rrbracket} - q_{M\,l}{}^{k\rrbracket}\e^l\right)
+2\,{\cal V}_{mn}{}^M\,\Omega_{pr} \Omega^{m\llbracket i} p_M{}^{jk\rrbracket np} \epsilon^{r}
+\frac12\,{\cal V}_{mn}{}^M\,p_M{}^{mn\llbracket ij} \epsilon^{k\rrbracket}
\nonumber\\
&&{}
+\frac18\,(1-6\lambda_\epsilon)\,\Gamma_{KM}{}^K\,{\cal V}^{\llbracket ij\,M} \epsilon^{k\rrbracket} 
\;,
\eea
where we have suppressed all $\omega_M$ contributions (which enter canonically),
and used (\ref{ident3}) to simplify the expression in the second line.

The algebra of the supersymmetry transformations closes on the $(1+4)$-dimensional general coordinate transformations (\ref{skewD}), 
generalized internal diffeomorphisms (\ref{gen_lie}), covariant gauge transformations of the $p$-form fields (\ref{deltaAB}), 
local ${\rm SO}(1,4)$ and ${\rm USp}(8)$ rotations, and an additional supersymmetry transformation, higher order in the fermions. 
The structural form of the supersymmetry algebra is the same as for the five-dimensional theory~\cite{deWit:2004nw}
\begin{equation}
\begin{aligned}
[\d(\e_1),\d(\e_2)]=\;&\x^\m\mc{D}_\m+\d_{\mathfrak{so}(1,4)}(\W^{ab})+\d_{\mathfrak{usp}(8)}(\L^{ij})+\d_{\rm susy}(\e_3)\\
& +\d_{\rm gauge}(\L^M)+\d_{\rm gauge}(\X_{\m\,M})+\d_{\rm gauge}(\X_{\m\n\,\bfa{}})+\d_{\rm gauge}(\Xi_{\m\n\,M})\;.
\end{aligned}
\label{susy_algebra}
\end{equation}
The transformation parameters on the r.h.s.\ can be explicitly given as combinations of the spinors $\e_{1,2}$, 
their covariant derivatives, and the external and internal vielbeins $e_\mu{}^a$, $\mc{V}_M{}^{ij}$, as
\bea
\x^\m&=&\fr12\bar{\e}_{2i}\g^\m\e_1^i\;,\label{susy_algebra2}\\
 \W^{ab}&=&-\fr{\sqrt{2}\,i}{3}\left(\bar{\e}_{1i}\g^{ab}{\nabla}^-_M\e_2^k- {\nabla}^-_M\bar{\e}_{1i}\g^{ab}\e_2^k\right)\mc{V}^{ij\, M}\W_{jk}-\L^M\w^-_M{}^{ab}\;,\nonumber\\
 \L^M&=&-\sqrt{2}i\,\mc{V}^{ij\,M}\W_{jk}\,\bar{\e}_{2i}\e^k_1\;,\nonumber\\
\X_{\m\,M}&=&\fr{1}{\sqrt{5}}\,\mc{V}_M{}^{kl}\W_{lm}\big(\bar{\e}_{2k}\g_\m\e_1^m\big)\;,\nonumber\\
\X{}_{\m\n\,\bfa}&=&\fr{3i}{\sqrt{10}}\,(t_\bfa)^M{}_N\mc{V}_{M\,li}\mc{V}^{ki\,N}\big(\bar{\e}_{2k}\g_{\m\n}\e_1^l\big)\;,\nonumber\\
\Xi_{\m\n\,M}&=&-\fr{i}{\sqrt{10}}\,\Big(\bar{\e}_{2k}\g_{\m\n}\dt_M\e_1^k-\dt_M\bar{\e}_{2k}\g_{\m\n}\e_1^k-(\bar{\e}_{2k}\e_1^k) \,e^a{}_{[\m}\dt_Me_{\n]a}-\fr23\, \mc{V}^{ki\,N}\dt_M \mc{V}_{N\,li}\big(\bar{\e}_{2k}\g_{\m\n}\e_1^l\big)\Big)\;.\nonumber
\eea

In the rest of this section we provide the explicit calculations that show closure of the supersymmetry algebra (\ref{susy_algebra}), (\ref{susy_algebra2}),
thereby confirming the supersymmetry transformation laws (\ref{transf}).
Let us start with closure on the external vielbein $e_\mu{}^a$
\bea
[\d_{\e_1},\d_{\e_2}]\,e_\m{}^a&=&\fr12\bar{\e}_{2i}\g^a\mc{D}_\m\e_1^i
- \fr{i}{\sqrt{2}}\,\bar{\e}_{2i}\mc{V}_M{}^{ij}\g^a\left({\nabla}^-_M(\g_\m\e_2^k)-\fr13 \g_\m{\nabla}^-_M\e_1^k\right)\W_{jk}-(1\leftrightarrow2)\nonumber\\
&=&\fr12\,\mc{D}_\m\left(\bar{\e}_{2i}\g^\n\e_1^ie_\n{}^a \right)- \sqrt{2}i\left(\bar{\e}_{2i}\e^k_1\mc{V}^{ij\,M}\W_{jk}\right){\nabla}^-_Me_\mu{}^a- \fr{\sqrt{2}i}{3}\,{\nabla}^-_M\left(\bar{\e}_{2i}\e^k_1\right)\mc{V}^{ij\,M}\W_{jk}e_\m{}^a\nonumber\\
&&{}-\fr{\sqrt{2}i}{3}\left(\bar{\e}_{2i}\g^{ab}{\nabla}^-_M\e_1^k- {\nabla}^-_M\bar{\e}_{2i}\g^{ab}\e_1^k\right)\mc{V}^{ij\,M}\W_{jk}e_{\m b}\;.
\eea
Taking into account that the term $\bar{\e}_{2i}\e^k_1$ has all spinor indices contracted, the generalized vielbein postulate (\ref{gvp}), and the vanishing torsion (\ref{torsion_ext}), we may rewrite the above expression as follows
\begin{equation}
\begin{aligned}
[\d_{\e_1},\d_{\e_2}]\,e_\m{}^a&=e_\n{}^a \mc{D}_\m \x^\n+\x^\n\mc{D}_\n e_\m{}^a+\L^N\dt_N e_\m{}^a+\fr13\,\dt_N\L^N e_\m{}^a+\W^{ab}e_{\m b}\;,
\end{aligned}
\label{dde}
\end{equation}
reproducing the correct transformation under external and internal diffeomorphisms. In particular, we obtain the correct value $\l=1/3$
for the weight of the f\"unfbein.

Next we check closure of the supersymmetry on the generalized vielbein $\mc{V}_M{}^{ij}$. We directly project the variation onto its coset valued part,
since any remaining part can be absorbed into a local ${\rm USp}(8)$ transformation.
The result is
\begin{equation}
\begin{aligned}
\mc{V}^{\llbracket kl\,M}\,[\d_{\e_1},\d_{\e_2}]\mc{V}_M{}^{ij\rrbracket}&=2\,{\cal P}_\m{}^{m\llbracket ijk}\W^{l\rrbracket p}\W_{mn}\big(\bar{\e}_{2p}\g^\m\e_1^n\big)
+6\sqrt{2}i\,\mc{V}^{\llbracket kl\,M}\W^{j|p|}\bar{\e}_{2p}{\nabla}^-_M\e_1^{i\rrbracket }-(1\leftrightarrow 2)\\
&=\x^\m {\cal P}_\m{}^{klij}+6\,\mc{V}^{\llbracket kl\,M}{\nabla}_M\big(\mc{V}_N{}^{ij\rrbracket }\L^N\big)\;,
\end{aligned}
\end{equation}
where we used the identity ${\cal P}_\m{}^{[ijkl}\W^{mn]}=0$ and the vielbein postulate. The first term in the expression above gives just a 
(covariantized) diffeomorphism along $\x^\m$, while the second can be rewritten using the generalised vanishing torsion condition (\ref{VT12}) which gives
\bea
\mc{V}^{\llbracket kl\,M}\,[\d_{\e_1},\d_{\e_2}]\mc{V}_M{}^{ij\rrbracket}& =&
\x^\m {\cal P}_\m{}^{klij}+6\,\mc{V}^{\llbracket kl\,M}\,\mc{D}_M\big(\mc{V}_N{}^{ij\rrbracket }\big)\L^N
+6\,\mc{V}^{\llbracket kl\,M}\,\mc{V}_N{}^{ij\rrbracket }\Big(\dt_M\L^N-\fr14\L^N\G_{KM}{}^K\Big)\nonumber\\
&=&\x^\m {\cal P}_\m{}^{klij}+\big(\mc{V}^{\llbracket kl\,M}\,\mc{D}_K\mc{V}_M{}^{ij\rrbracket }\big)\L^N+6\,\mc{V}^{\llbracket kl\,M}\,\mc{V}_N{}^{ij\rrbracket }\,\dt_M\L^N\nonumber\\
&=&\x^\m {\cal P}_\m{}^{klij}+\mc{V}^{\llbracket kl\,M}\,\d_\L\mc{V}_M{}^{ij\rrbracket}\;.
\eea
The weight term that comes from the derivative of $\mc{V}_N{}^{ij}\L^N$ is cancelled by the same contribution from the vanishing torsion condition. 
Again, we find the correct transformation with the same gauge parameters as in (\ref{dde}).

Now we turn to the gauge field sector and investigate closure of the supersymmetry algebra on the vector field ${\cal A}_\m{}^M$.
A direct calculation gives
\begin{equation}
\begin{aligned}
[\d_{\e_1},\d_{\e_2}]\,{\cal A}_\m{}^M&=\sqrt{2}i\,\W^{ik}(\bar{\e}_{2k}\mc{D}_\m\e_1^j)\mc{V}_{ij}{}^M
+\sqrt{2}i\,\bar{\e}_{2k}\g_\m {\cal P}_\n{}^{ijkl}\g^\n\W_{lm}\e_1^m\mc{V}_{ij}{}^M\\
&\quad +3\,\bar{\e}_{2k}\g_\m\left(\W^{m[i}\mc{V}^{jk]N} -\fr13\,\W^{[ij}\mc{V}^{k]m\,N} \right)\W_{mr}{\nabla}^-_N\e^r_1\mc{V}_{ij}{}^M\\
&\quad+2\,\W^{ik}\bar{\e}_{2k}\mc{V}^{jr\,N}\left({\nabla}^-_N(\g_\m\e_1^s)-\fr13\g_\m{\nabla}^-_N\e_1^s\right)\W_{rs}\mc{V}_{ij}{}^M- (1\leftrightarrow 2)\\
&=\mc{D}_\m\L^M+\fr12\dt_N\big(\bar{\e}_{2k}\g_\m\e_1^k\big)\mc{M}^{MN}-\big(\bar{\e}_{2k}\g^\n\e_1^k\big)\mc{M}^{MN}e_{\n a}{\nabla}^-_Ne_\m{}^a\\
&\quad+2\Big(\mc{V}^{Mik}\mc{V}_{ij}{}^N+\mc{V}^{Nik}\mc{V}_{ij}{}^M-\fr14\d^k_j\mc{M}^{MN}\Big){\nabla}^-_N\big(\bar{\e}_{2k}\g_\m\e_1^j\big)\\
&=\mc{D}_\m\L^M+g_{\m\n}\dt_N\x^\n\mc{M}^{MN}-2\x^\n\mc{M}^{MN}e_{a[\n}{\nabla}^-_Ne_{\m]}{}^a\\
&\quad-2\,{\nabla}^-_N\bigg[\Big(\mc{V}^{ik\,M}\mc{V}_{ij}{}^N+\mc{V}^{ik\,N}\mc{V}_{ij}{}^M-\fr14\d^k_j\mc{M}^{MN}\Big)\big(\bar{\e}_{2k}\g_\m\e_1^j\big)\bigg]\;.
\end{aligned}
\end{equation}
Finally using the relations \eqref{d2W} and (\ref{omega_int}), the above expression can be written in the following form
\begin{equation}
\begin{aligned}
[\d_{\e_1},\d_{\e_2}]\,{\cal A}_\m{}^M&=g_{\m\n}\dt_N\x^\n\mc{M}^{MN}-2\x^\n\mc{M}^{MN}e_{a[\n}{\nabla}^-_Ne_{\m]}{}^a\\
&\quad+\mc{D}_\m\L^M-10\,d^{MNK}\dt_N\X_{\m\,K}\\[1ex]
&=\x^\n\F_{\n\m}{}^M+g_{\m\n}\dt_N\x^\n\mc{M}^{MN}+\mc{D}_\m\L^M-10\,d^{MNK}\dt_N\X_{\m\,K},
\end{aligned}
\end{equation}
with the parameter $\Xi_{\mu\,M}$ from (\ref{susy_algebra2}), thus precisely reproducing 
the ${\rm E}_{6(6)}$ covariant gauge transformation (\ref{deltaAB}) of the gauge field coming from tensor hierarchy.
The first two terms in the expression correspond to the transformation (\ref{skewD}) of the gauge field under external diffeomorphisms.

Finally, we investigate transformations of the two-form field ${\cal B}_{\m\n\, M}$ that give
\bea
\label{clos_B}
[\d_{\e_1},\d_{\e_2}]\,{\cal B}_{\m\n\,M} &=&
-\fr{1}{\sqrt{5}}\,
\Big[2\,\mc{V}_M{}^{ij}\bar{\e}_{2i}\g_{[\m}\mc{D}_{\n]}\e_1^k\W_{jk} +\fr12\,\mc{V}_{M\,lm}\bar{\e}_{2n}\g_{\m\n}\g^\r \mc{P}_\r{}^{lmnp}\W_{pq}\e_1^q\Big]\nonumber\\
&&{}-\fr{4i}{\sqrt{10}}\,\mc{V}_M{}^{ij}\mc{V}_{jp}{}^{N}\Big(\bar{\e}_{2i}\g_{[\m}{\nabla}^-_N(\g_{\n]}\e_1^p)-\fr13\bar{\e}_{2i}\g_{\m\n}{\nabla}^-_N\e_1^p\Big)\nonumber\\
&&{}-\fr{3i}{\sqrt{10}}\mc{V}_{M\,lm}\bar{\e}_{2n}\g_{\m\n}\Big(\W^{k[l}\mc{V}^{mn]N}-\fr13\W^{[lm}\mc{V}^{n]k\,N}\Big)\W_{kp}{\nabla}^-_N\e_1^p
~-~(1\leftrightarrow 2)\nonumber\\
&=&\fr{1}{\sqrt{5}}\,\mc{D}_{[\m}\Big(\mc{V}_M{}^{ij}\bar{\e}_{2i}\g_{\n]}\e_1^k\,\W_{jk}\Big)
-\fr{4i}{\sqrt{10}} \,\mc{V}_{jl}{}^N\mc{V}^{kj\,M}\big(\bar{\e}_{2k}\g_{a}\g_b\e_1^l\big)e_{[\m}{}^a {\nabla}^-_Ne_{\n]}{}^b\nonumber\\
&&{}-\fr{2i}{\sqrt{10}}\,\Big(\mc{V}_{M\,ni}\mc{V}^{ki\,N}-\mc{V}_{ni}{}^N\mc{V}_{M}{}^{ki}\Big)\big(\bar{\e}_{2k}\g_{\m\n}{\nabla}^-_N\e_1^n\big)
+\fr{i}{\sqrt{10}}\, \big(\bar{\e}_{2k}\g_{\m\n}{\nabla}^-_M\e_1^k\big)
\nonumber\\
&&{}~-~(1\leftrightarrow 2)\nonumber\\
&=&2\,\mc{D}_{[\m}\X_{\n]M}-\fr{2i}{\sqrt{10}}\,\Big(\mc{V}_{M\,ni}\mc{V}^{ki\,N}-\mc{V}_{ni}{}^N\mc{V}_{M}{}^{ki}\Big)\,{\nabla}^-_N\big(\bar{\e}_{2k}\g_{\m\n}\e_1^n\big)\nonumber\\
&&{}-\fr{4i}{\sqrt{10}}\,\Big(\mc{V}_{M\,ni}\mc{V}^{ki\,N}-\mc{V}_{ni}{}^N\mc{V}_{M}{}^{ki}\Big)(\bar{\e}_{2k}\e_1^n) \,e{}_{[\m}{}^a{\nabla}^-_Ne_{\n]a}\nonumber\\
&&{}+\fr{i}{\sqrt{10}}\,\Big(\bar{\e}_{2k}\g_{\m\n}{\nabla}^-_M\e_1^k-{\nabla}^-_M\bar{\e}_{2k}\g_{\m\n}\e_1^k\Big)\;.
\eea
Here, we have systematically ignored the contribution from the last term $-d_{MNP}{\cal A}_{[\mu}{}^N \delta {\cal A}_{\nu]}{}^P$ in the supersymmetry
variation, which will simply reproduce the corresponding terms in the action of generalized diffemorphisms and gauge transformations, due to the fact 
that the algebra closes on the vector field ${\cal A}_{\mu}{}^M$.
To simplify the second term in (\ref{clos_B}) it is helpful to consider the following identity
\begin{equation}
\begin{aligned}
\Big(\mc{V}_{M\,ni}\mc{V}^{ki\,N}-\mc{V}_{ni}{}^N\mc{V}_{M}{}^{ki}\Big)\big(\bar{\e}_{2k}\g_{\m\n}\e_1^n\big)
&=\Big(\mc{V}_{M\,ni}\mc{V}^{ki\,N}\W_{mk}
+\mc{V}_{M\,mi}\mc{V}^{ki\,N}\W_{nk}\Big)\W^{mp}\big(\bar{\e}_{2p}\g_{\m\n}\e_1^n\big)\\
&=\fr32\,\mathbb{P}_M{}^N{}_Q{}^P\mc{V}_{P\,ni}\mc{V}^{ki\,Q}\big(\bar{\e}_{2k}\g_{\m\n}\e_1^n\big)\;,
\end{aligned}
\end{equation}
where in the second line we notice that the expression in brackets is symmetric in $(mn)$ and hence is an element of the $\mathfrak{usp}(8)$ part of $\mathfrak{e}_{6(6)}$. The traceless antisymmetrisation on the r.h.s.\ can be replaced by the usual antisymmetrization giving the same result. Using this relation and the vanishing torsion condition we may express the corresponding term as follows
\bea
\lefteqn{{\nabla}_N\bigg[\Big(\mc{V}_{M\,ni}\mc{V}^{ki\,N}-\mc{V}_{ni}{}^N\mc{V}_{M}{}^{ki}\Big)\big(\bar{\e}_{2k}\g_{\m\n}\e_1^n\big)\bigg]~=}
\nonumber\\
&=&\dt_N\bigg[\Big(\mc{V}_{M\,ni}\mc{V}^{ki\,Nki}-\mc{V}_{ni}{}^N\mc{V}_{M}{}^{ki}\Big)\big(\bar{\e}_{2k}\g_{\m\n}\e_1^n\big)\bigg]-\fr13\mc{V}_{N\,ni}\mc{D}_M\mc{V}^{ki\,N}\big(\bar{\e}_{2k}\g_{\m\n}\e_1^n\big)\nonumber\\
&=&\fr{\sqrt{10}\,i}{2}\,(t_\bfa){}_M{}^N\,\dt_N\X_{\m\n}{}^\bfa - \fr13\mc{V}_{N\,ni}\mc{D}_M\mc{V}^{ki\,N}\big(\bar{\e}_{2k}\g_{\m\n}\e_1^n\big)\;,
\label{ksi_a}
\eea
with $\X_{\m\n}{}^\bfa$ from (\ref{susy_algebra2}). 
Next, using the identity \eqref{d2W}, the second line of the last equation in \eqref{clos_B} can be rewritten in the following suggestive form
\begin{multline}
\label{d_H}
-\fr{4i}{\sqrt{10}}\Big(\mc{V}_{M\,ni}\mc{V}^{ki\,N}+\mc{V}_{ni}{}^N\mc{V}_{M}{}^{ki}\Big)(\bar{\e}_{2k}\e_1^n) \,e{}_{[\m}{}^a {\nabla}^-_Ne_{\n]a}\\
= d_{MNK}\L^N\mc{F}{}_{\m\n}{}^K-\fr{i}{\sqrt{10}}\,(\bar{\e}_{2k}\e_1^k) \,e{}_{[\m}{}^a {\nabla}^-_M e_{\n]a}.
\end{multline}
Finally, we focus on the last term in the last equation of \eqref{clos_B} and notice that its ${\rm USp}(8)$ connection part cancels that of the last term in \eqref{ksi_a}. Hence, we may take into account only the spin connection ${\w}^-_M{}^{\r\s}$ that includes the $SO(1,4)$ connection and the field strength $\mc{F}^{\r\s\,M}$. After some gamma-matrices algebra we obtain the following expression
\begin{equation}
\begin{aligned}
\bar{\e}_{2k}\g_{\m\n}{\nabla}^-_M\e_1^k-{\nabla}^-_M\bar{\e}_{2k}\g_{\m\n}\e_1^k&=-\fr{1}{4}\,\mc{M}_{MN}\big(\bar{\e}_{2k}\g_{\m\n\r\s}\e_1^k\big)\mc{F}^{\r\s N}+(\bar{\e}_{2k}\e_1^k) e{}_{[\m}{}^a{\nabla}^-_Me_{\n]a}\\
&\quad+\bar{\e}_{2k}\g_{\m\n}\dt_M\e_1^k-\dt_M\bar{\e}_{2k}\g_{\m\n}\e_1^k-(\bar{\e}_{2k}\e_1^k) e{}_{[\m}{}^a\dt_Me_{\n]a}\\
&=-\fr{i}{2} \,\x^\l e\ve_{\l\m\n\r\s}\,\mc{M}_{MN}\mc{F}^{\r\s N}+(\bar{\e}_{2k}\e_1^k) e_{[\m}{}^a{\nabla}^-_Me_{\n]a}\\
&\quad+\bar{\e}_{2k}\g_{\m\n}\dt_M\e_1^k-\dt_M\bar{\e}_{2k}\g_{\m\n}\e_1^k-(\bar{\e}_{2k}\e_1^k) e_{[\m}{}^a \dt_Me_{\n]a}.
\end{aligned}
\end{equation}
The first term here represents the diffeomorphism transformation (\ref{skewD}) of the field ${\cal B}_{\m\n}$, the second term precisely cancels the last term in \eqref{d_H}. The rest can be packaged into a tensor $\mc{O}_{M\m\n}$ constrained by
\begin{equation}
d^{MNK}\dt_N\mc{O}_{K\m\n}=0\;,
\end{equation}
as a consequence of the section condition. %
Collecting everything together, the commutator of supersymmetry transformations \eqref{clos_B} of the two-form field takes the following form
\begin{equation}
\begin{aligned}
[\d_{\e_1},\d_{\e_2}]\,{\cal B}_{\m\n\,M}&=2\,\mc{D}_{[\m}\X_{\n]M}+\fr{1}{2\sqrt{10}}\,\x^\l e\varepsilon_{\l\m\n\r\s}\mc{M}_{MN}\mc{F}^{\r\s N} 
+(t_\bfa)_M{}^N\,\dt_N\X{}_{\m\n}{}^\bfa\\
&\quad
+d_{MNK}\L^N\mc{F}^K{}_{\m\n}+\mc{O}_{M\m\n}
 -d_{MKL}\,{\cal A}_{[\mu}{}^K\, [\d_{\e_1},\d_{\e_2}]\, {\cal A}_{\nu]}{}^L\;,
\end{aligned} 
\end{equation}
up to terms of higher order in the fermions. This confirms the supersymmetry algebra (\ref{susy_algebra}), (\ref{susy_algebra2}).

\section{Invariant Lagrangian}
\label{invL}

We now have all the ingredients to present the full supersymmetric Lagrangian for ${\rm E}_{6(6)}$ exceptional field theory.
Its bosonic part has been constructed in~\cite{Hohm:2013pua,Hohm:2013vpa}, here we give the supersymmetric
extension based on the fermionic structures introduced in the previous sections.
The final result reads
\bea
\label{L} 
e^{-1}\mc{L}&=&\ \widehat{\cal R}-
\fr{1}{4}\,\mc{M}_{MN}\,\F_{\m\n}{}^{M}\F^{\m\n
N}-\fr{1}{6}\,\mc{P}_\m{}^{ijkl}\mc{P}^\m{}_{ijkl}+\frac{\sqrt{10}}{8}\,e^{-1}\,{\cal L}_{\rm top} - V(\mc{M},g)\nonumber\\ 
&&{}-\bar{\y}_{\m i}\g^{\m\n\r}\mc{D}_\n\y_\r^i+2\sqrt{2}\,i\,\mc{V}_{ij}{}^M\W^{ik}\bar{\y}_{\m k}\g^{[\m}{{\nabla}}^+_M\left(\g^{\n]}\y_\n{}^j\right)\nonumber\\ 
&&{}-\fr{4}{3}\,\bar{\c}_{ijk}\g^\m\mc{D}_\m\c^{ijk}+8\sqrt{2}\,i\,\mc{V}_{mn}{}^M\W^{np}\bar{\c}_{pkl} {{\nabla}}^+_M\c^{mkl}\nonumber\\ 
&&{}+\fr{4i}{3}\mc{P}_\m{}^{ijkl}\bar{\c}_{ijk}\g^\n\g^\m\y_\n{}^m\W_{lm} +4\sqrt{2}\,\mc{V}^{ij\,M}\,\bar{\c}_{ijk}\g^\m{{\nabla}}^-_M\y_\m{}^k\;,
\eea
up to quartic fermion terms. The latter are expected to coincide with the quartic terms of the $D=5$ theory~\cite{Cremmer:1980gs}.
Let us explain the various terms of (\ref{L}).
The first line describes the bosonic couplings, with the modified Ricci scalar 
$\widehat{{\cal R}}$ obtained from contracting (\ref{RiemannHat}), Yang-Mills term for the field strength (\ref{defFH}) and the scalar kinetic term
\bea
-\fr{1}{6}\,\mc{P}_\m{}^{ijkl}\mc{P}^\m{}_{ijkl} &=& \frac1{24}\,{\cal D}_\mu {\cal M}_{MN} {\cal D}^\mu {\cal M}^{MN}
\;.
\eea
We note, that variation of the Einstein-Hilbert term and the scalar kinetic term w.r.t.\ the vector fields is given by
\bea
\delta \left( e\,\widehat{\cal R}-\fr{1}{6}\,e\,\mc{P}_\m{}^{ijkl}\mc{P}^\m{}_{ijkl}  \right) &=& 
e\left(\widehat{J}^\mu{}_{M} +{\cal J}^\mu{}_{M} \right) \delta {\cal A}_\mu{}^M\;,
\label{varMin}
\eea
with the current $\widehat{J}^\mu{}_{M}$ from (\ref{STep}) and the scalar current given by
\bea 
\label{var_EH_A}
{\cal J}^\mu{}_{M} &=&
 -2\,{\cal V}_M{}_{ij}\,{\cal V}_{kl}{}^{N} \,\nabla_N\left( g^{\mu\nu}  {\cal P}_{\nu}{}^{ijkl} \right)
 \;.
 \label{JMM}
\eea

The topological term in (\ref{L}) is most compactly defined by its variation
\bea
\delta {\cal L}_{\rm top}&=& 
\varepsilon^{\mu\nu\rho\sigma\tau}
\Big(d_{MNK}
{\cal F}_{\mu\nu}{}^M {\cal F}_{\rho\sigma}{}^N  \delta {\cal A}_\tau{}^K
+\frac{20}{3}\,d^{MNK}\partial_N{\cal H}_{\mu\nu\rho \,M} \,
\left(\delta {\cal B}_{\sigma\tau\,K}+d_{KPQ}{\cal A}_{\sigma}{}^P\d  {\cal A}_{\tau}{}^Q\right) \Big)
\;,\nonumber\\
\label{vartop}
\eea
equivalently, the associated action can be expressed as the boundary contribution of a manifestly covariant integral 
over six external dimensions. The scalar potential $V$ has been given in~\cite{Hohm:2013pua,Hohm:2013vpa} in 
the explicit form
\begin{equation}
\begin{aligned}
  V(\mc{M},g)=& -\frac{1}{24}\,{\cal M}^{MN}\partial_M{\cal M}^{KL}\,\partial_N{\cal M}_{KL}
  +\frac{1}{2} \,{\cal M}^{MN}\partial_M{\cal M}^{KL}\partial_L{\cal M}_{NK}\\
  &{}-\frac{1}{2}\,g^{-1}\partial_Mg\,\partial_N{\cal M}^{MN}
  -\frac{1}{4}  \,{\cal M}^{MN}g^{-1}\partial_Mg\,g^{-1}\partial_Ng
  -\frac{1}{4}\,{\cal M}^{MN}\partial_Mg^{\mu\nu}\partial_N g_{\mu\nu}\;,
\end{aligned}
\label{Vpot}
\end{equation}
and can be rewritten in the following manifestly covariant form 
\begin{equation}\begin{aligned}
V(\mc{M},g)&= {\cal R}  -\frac{1}{4}\, {\cal M}^{MN}\,\nabla_Mg_{\mu\nu}\nabla_N g^{\mu\nu} + \nabla_M I^M
\;,
\label{Vcov}
\end{aligned}\end{equation}
with the curvature scalar ${\cal R}$ from (\ref{Rijkl}), up to boundary contributions $I^M$
and terms that vanish due to the section condition.
The explicit calculation confirming (\ref{Vcov}) requires a number of non-trivial ${\rm USp}(8)$ identities, some
of which are collected in appendix~\ref{app:identities}.

The kinetic fermion terms in (\ref{L}) are such that 
upon dropping all internal derivatives, the Lagrangian ${\cal L}_0\equiv{\cal L}|_{\partial_M\rightarrow0}$ 
reduces to the five-dimensional theory~\cite{Cremmer:1980gs,deWit:2004nw}.
The fermion terms carrying internal derivatives $\nabla_M$ are then obtained by imposing invariance of the Lagrangian
under the supersymmetry transformations (\ref{transf}).\footnote{See also~\cite{Coimbra:2012af} for these couplings in a Cliff$(10,1;\mathbb{R})$ formulation.}
In the limit $\partial_M\rightarrow0$, these terms reduce to the Pauli couplings of fermions to the field strength via (\ref{omegaHat}) and again
reproduce the couplings from the $D=5$ theory. It is interesting to observe that in the full theory, 
and unlike for the supersymmetry transformations (\ref{transf}), these ${\cal F}_{\mu\nu}{}^M$
couplings cannot entirely be absorbed into a homogeneous shift of the internal spin connection (\ref{omegaHat}), 
but require both $\nabla_M^+$ and $\nabla_M^-$ derivatives, however in a very systematic pattern.

By construction, the full Lagrangian (\ref{L}) is manifestly invariant under generalized internal diffeomorphisms. 
To show that it is invariant under supersymmetry, one has to go through rather tedious calculations, 
that we sketch in the remainder of this section. For the full detailed calculations the reader is referred to Appendix~\ref{app:details}.
The proof of supersymmetry of the Lagrangian is most conveniently organized order by order in the 
internal derivatives $\nabla_M$.\footnote{This is very much in parallel with the analogous calculation in gauged 
supergravity~\cite{deWit:2004nw} order by order in the coupling constant.}
Internal derivatives enter in ${\cal L}$ in two different ways: first they render the Lagrangian ${\cal L}_0$ 
covariant under generalized diffeomorphisms by virtue of (\ref{covD}) and (\ref{defFH}), 
second they give rise to explicit couplings such as
the bilinear fermion terms and the scalar potential $V$. I.e.\ the Lagrangian schematically organizes as
\bea
{\cal L}&=& {\cal L}_0^{\rm cov} + {\cal L}_1[\bar\psi \nabla_M \psi] + {\cal L}_2[\nabla_M{\cal M}\nabla_N{\cal M}]
\;.
\eea
Similarly, the supersymmetry transformations (\ref{transf}) organize as
\bea
\delta &=& \delta_0^{\rm cov} + \delta_1[\nabla_M\epsilon]
\;,
\eea
where $\delta_0$ describe the supersymmetry transformation laws of the five-dimensional theory.
In lowest order in $\nabla_M$, supersymmetry of the Lagrangian  
amounts to the corresponding property of the five-dimensional theory~\cite{Cremmer:1980gs,deWit:2004nw}. 
In first and second order in $\nabla_M$, the contributions from $\delta_0^{\rm cov} {\cal L}_1$,
$\delta_1 {\cal L}_0^{\rm cov}$, and $\delta_1 {\cal L}_1$ can be organized according to their 
fermion structure
\begin{equation}
\begin{aligned}
& \y\mc{D}_\m \nabla_M \e, && \c \mc{D}_\m \nabla_M \e, &&
\y \nabla_M \nabla_N \e, && \c \nabla_M\nabla_N \e\;,
\end{aligned}
\label{4terms}
\end{equation}
and we discuss the four classes of terms separately in appendices \ref{subsec:DdPsi} -- \ref{subsec:DDChi}.
The latter terms combine with the second order contributions from $\delta_0^{\rm cov} {\cal L}_2$ arising from
variation of the scalar potential (\ref{Vcov}). These are obtained by using the properties of the scalar curvature (\ref{varR1}), (\ref{varR2}) as
\bea
\delta_\epsilon (eV) &=&
\frac12\,e\left(\,g^{\mu\nu}\, {\cal R} 
 -\frac{1}{4}\,g^{\mu\nu}\,{\cal M}^{MN}\nabla_Mg^{\mu\nu}\nabla_N g_{\mu\nu}
  +\nabla_N ({\cal M}^{MN}\nabla_Mg^{\mu\nu}) \right) \delta_\epsilon g_{\mu\nu}
   \nonumber\\
 &&{}
+ e\,\Sigma^\epsilon_{ijkl}\left( {\cal R}^{ijkl}
 -\frac{1}{2}\, {\cal V}^{ij\,M} {\cal V}^{kl\,N}\,\nabla_Mg_{\mu\nu}\nabla_N g^{\mu\nu} \right)
 \;,
 \label{varV}
\eea
up to total derivatives, and
with $\Sigma^\epsilon_{ijkl}\equiv -4i\,\W_{m\llbracket i}\bar{\c}_{jkl\rrbracket}\e^m$ describing the supersymmetry variation
of the scalar fields (\ref{transf}).

In addition, we have further contributions from $\delta_0^{\rm cov} {\cal L}_0^{\rm cov}$ 
due to the fact that covariant derivatives ${\cal D}_\mu$ no longer commute.
Such contributions arise from variation of the fermionic kinetic term with (\ref{DPcov})
\bea
-\fr{4i}{3}\,\bar{\c}_{ijk}\g^{\m\n} \epsilon^m\, \mc{D}_\m {\cal P}_{\nu}{}^{ijkl} \Omega_{lm} 
&=&
4i\,\bar{\c}_{ijk}\g^{\m\n} \epsilon^m \Omega_{lm} \, {\cal V}_N{}^{\llbracket ij}{\cal V}^{kl\rrbracket M} \,\nabla_M\F_{\mu\nu}{}^N
\;,
\label{varF}
\eea
but also from variation of the Rarita-Schwinger term upon using the commutator (\ref{comm_ext})
\bea
-\bar\psi_{\mu\,i} \gamma^{\mu\nu\rho} \left[{\cal D}_\nu, {\cal D}_\rho\right] \epsilon^i &=&
- \bar\psi_{\mu\,i}\gamma_\nu \epsilon^i  \,\left(\widehat{\cal R}^{\nu\mu} -\frac12\,g^{\mu\nu}\,\widehat{\cal R}\right) 
-\fr23\,{\cal P}_{\nu}{}^{iklm}\,{\cal P}_{\rho\, jklm}\,\bar\psi_{\mu\,i} \gamma^{\mu\nu\rho} \epsilon^j
\nonumber\\
&&{}
+{\cal F}_{\nu\rho}{}^M\,\bar\psi_{\mu\,i} \gamma^{\mu\nu\rho} \nabla_M \epsilon^i
- \nabla_{M} \F_{\nu\rho}{}^N  \left(
{\cal V}_N{}^{jk} {\cal V}_{ik}{}^M-
 {\cal V}_{N\,ik} {\cal V}^{jk\,M} \right)\bar\psi_{\mu\,i} \gamma^{\mu\nu\rho} \epsilon^j 
\nonumber\\
&&{}
+\frac12\, \bar\psi_{\mu\,i} \gamma^{\mu\rho}{}_\nu \epsilon^i \, \nabla_M {\cal F}_{\rho}{}^{\nu\,M}
- \frac14\,\bar\psi_{\mu\,i} \gamma^{\nu\rho\sigma} \epsilon^i \,{\cal F}_{\nu\rho}{}^K g^{\mu\tau} \nabla_K g_{\sigma\tau} \,
\;.
\label{varRS}
\eea
Here the first two terms cancel as in the $D=5$ theory (where it is important though that $\widehat{\cal R}^{\mu\nu}$
arises with indices contracted in the proper order since $\widehat{\cal R}^{[\mu\nu]}\not=0$), while all remaining terms cancel
against terms of the form (\ref{4terms}) as discussed in appendix~\ref{subsec:DDPsi}, \ref{subsec:DDChi}.

Finally, there are the contributions that arise from variation of the vector gauge field in the minimal couplings of (\ref{covD}),
and from variation of the two-form gauge field in the vector kinetic term and the topological term. The first appear proportional
to the currents from (\ref{varMin})
\bea
\delta_A \left( e\widehat{\cal R}-\fr{1}{6}\,e\,\mc{P}_\m{}^{ijkl}\mc{P}^\m{}_{ijkl}  \right) &=& 
e\left(\widehat{J}^\mu{}_{M} +{\cal J}^\mu{}_{M} \right) \delta {\cal A}_\mu{}^M\;,
\label{varAA}
\eea
and the latter are proportional to the first order duality equation between vectors and tensors
\bea
\delta_B {\cal L} &=& 
5\,d^{MNK} \nabla_N
\Big(e{\cal M}_{MN} {\cal F}^{\mu\nu\,N} 
+
\frac{\sqrt{10}}{6}\, \varepsilon^{\mu\nu\rho\sigma\tau}
\,{\cal H}_{\rho\sigma\tau \,M} \Big)
\left(\delta_\epsilon {\cal B}_{\mu\nu\,K}+d_{KPQ}{\cal A}_{\mu}{}^P\d_\epsilon  {\cal A}_{\nu}{}^Q\right) 
\;.\nonumber\\
\label{varBB}
\eea
All these terms cancel against terms of the form (\ref{4terms}) as discussed in appendix~\ref{subsec:DdPsi}--\ref{subsec:DDChi}.

As a final result, we find that the Lagrangian (\ref{L}) is supersymmetric under the transformations (\ref{transf}) up to terms of
higher order in the fermions. Remarkably, and unlike in the reduced theory, invariance of the Lagrangian under generalized diffeomorphisms
(\ref{gen_lie}), (\ref{skewD}) already fixes all the bosonic couplings without reference to supersymmetry.
The present construction gives the fermionic completion which turns the bosonic Lagrangian of~\cite{Hohm:2013pua,Hohm:2013vpa} into
a supersymmetric system.

\section{Conclusions and discussion}
\label{sec:conclusions}

In this paper we have constructed the supersymmetric completion of ${\rm E}_{6(6)}$-covariant exceptional field theory,
with the final result given by the Lagrangian (\ref{L}) and the supersymmetry transformation laws (\ref{transf}).
The section condition (\ref{sectioncondition}) effectively constrains the geometry of the extended space. 
It admits at least two independent maximal solutions which restrict the number of internal coordinates
to six and five, respectively~\cite{Hohm:2013pua,Hohm:2013vpa}.\footnote{
The same is true for the ${\rm E}_{7,8}$ cases and the higher dimensional ${\rm SO}(5,5)$ and ${\rm SL}(5)$ EFT's 
\cite{Hohm:2013uia,Blair:2013gqa,Hohm:2014fxa,Abzalov:2015ab}.}
They are identified upon splitting the $\bf 27$ representation of ${\rm E}_{6(6)}$ under the action of the subgroup 
${\rm GL}(6)$ and ${\rm GL}(5)\times {\rm SL}(2)$, respectively. 
Upon imposing the former solution, the Lagrangian (\ref{L}) reproduces the full Lagrangian of $D=11$ supergravity, as explicitly demonstrated for its
bosonic part in \cite{Hohm:2013vpa}. With the latter solution, the Lagrangian (\ref{L}) describes the full supersymmetric IIB theory.
It may at first appear surprising that one and the same set of fermions and couplings encodes both type IIA and type IIB, despite the crucial difference of their fermion chiralities.   
This is due to the fact that the ${\rm E}_{6(6)}$-covariant formulation (\ref{L}) does not preserve the original $D=10$ Lorentz invariance. As a consequence, its fermions can consistently encode the fermions of the type IIA and type IIB theory in the same way that both type IIA and type IIB give rise to the same supersymmetric  theory in $D=5$ upon dimensional reduction. 

Upon the most straightforward solution of the section constraint, that is $\dt_M=0$, the Lagrangian (\ref{L}) directly reduces to the 
maximal $D=5$ supergravity of~\cite{Cremmer:1980gs}. In the context of generalized Scherk-Schwarz reductions, 
it has been proposed to relax the section condition~(\ref{sectioncondition}) from a differential constraint into the known algebraic constraints on the embedding tensor, 
that naturally appears as a generalized torsion~\cite{Grana:2012rr,Berman:2012uy,Musaev:2013rq,Aldazabal:2013mya}. 
Although, the generalized torsion formally reproduces all the gaugings, it remains an open question, to which extent they can be embedded into higher-dimensional
supergravity via the corresponding EFT. The work \cite{Dibitetto:2012rk}, where the structure of the space of T-duality orbits was analysed, suggests that 
in principle one should be able to catch non-geometric compactifications by generalized Scherk-Schwarz reductions of EFT.
On the other hand, a generalized Scherk-Schwarz ansatz that is consistent with the section condition~(\ref{sectioncondition}), 
describes a consistent truncation of the exceptional field theory (\ref{L}) and by virtue of the section condition translates into a consistent truncation 
of the conventional higher-dimensional supergravities. For the
${\rm SO}(p,q)$ gauged supergravities, this ansatz has been constructed in~\cite{Hohm:2014qga}. It yields their higher-dimensional embedding 
as sphere and hyperboloid compactifications of the higher-dimensional supergravities~\cite{Hohm:2014qga},\footnote{See also~\cite{Baron:2014bya} 
for the explicit uplift of several vacua of these theories.} and naturally extends to the full Lagrangian (\ref{L}).

In discussion of geometry of the extended space let us mention the works \cite{Berman:2014jba,Cederwall:2014kxa,Cederwall:2014opa} where the geometrical meaning of the T--duality group ${\rm O}(d,d)$ has been investigated. It was conjectured that the $d$-dimensional torus is just one of possible solutions of the field equations of double field theory, precisely the one that preserves the whole ${\rm O}(d,d)$ group. Following this direction one may try to construct other solutions of DFT or EFT that preserve less duality symmetries and compare these with the known examples.
Recently, in~\cite{Berman:2014hna} it was shown that the brane solutions of $D=4$ supergravity can be uplifted to a single solution of ${\rm E}_{7(7)}$ exceptional field theory, that solves the twisted self-duality constraint. A possible direction of further research would be the investigation of similar uplifts in the presented ${\rm E}_{6(6)}$ theory adding, possibly, winding coordinates, that should lead to non-geometric branes. Following the lines of \cite{Godazgar:2014nqa} and the result of this paper one may explicitly investigate the supersymmetry properties of the obtained solutions in the EFT sense. In this context, we also mention the recent~\cite{Coimbra:2014uxa} for the embedding of supersymmetric flux backgrounds in exceptional geometry.

\section*{Acknowledgements}

On its last stage the work of ETM was partially supported by the German Science Foundation (DFG) under the Collaborative Research Center (SFB) 676 Particles, Strings and the Early Universe. ETM would like to thank Theoretical Department of CERN, Theory Department of DESY and II. Institut f\"ur Theoretische Physik, and personally Jan Louis, for warm hospitality during completion of part of this work. We thank Arnaud Baguet, Olaf Hohm, Hermann Nicolai, and Oscar Varela for inspiring discussions.

\section*{Appendix}
\begin{appendix}

\section{Details of the supersymmetry calculation}
\label{app:details}

In this section we provide most of the technical details of the rather lengthy calculations required to verify supersymmetry invariance of the Lagrangian \eqref{L} under the transformations (\ref{transf}).  We discuss the various cancellations according to the different types of terms (\ref{4terms}) that arise in the variation of the Lagrangian.

\subsection{The  \texorpdfstring{$\y \nabla_M \mc{D}_\m  \e$}{PsiNablaDe} terms}
\label{subsec:DdPsi}

The relevant contributions of this type from variation of the Rarita-Schwinger term are
\bea
\d_\e(-e\bar{\y}_{\m i}\g^{\m\n\r}\mc{D}_\n\y_\r^i)&\longrightarrow&-\bar{\y}_{\m i}\mc{D}_\n(e\g^{\m\n\r})\d_\e\y_\r^i
-2e\bar{\y}_{\m i}\g^{\m\n\r}\mc{D}_\n\d_\e\y_\r^i\\
&\longrightarrow& -2\sqrt{2}i e\mc{D}_\n\bar{\y}_{\m
i}\g^{\m\n\r}\left(\fr23\g_\r {\nabla}^-_M\e^k\mc{V}^{ij\,M}\W_{jk}+{
\nabla}^-_M\g_\r \e^k\mc{V}^{ij\,M}\W_{jk}\right)\nonumber\\
&=&-4\sqrt{2}i e\mc{D}_\n\bar{\y}_{\m
i}\g^{\m\n}{\nabla}^-_M\e^k\mc{V}^{ij\,M}\W_{jk}-2\sqrt{2}i
e\mc{D}_\n\bar{\y}_{\m i}\g^{\m\n\r}{\nabla}^-_M\g_\r
\e^k\mc{V}^{ij\,M}\W_{jk}\;,\nonumber
\eea
where the term $\mc{D}_\n(e\g^{\m\n\r})$ vanishes due to the vanishing torsion condition. The other contributions of this type come from the following variations
\begin{equation}
\begin{aligned}
\d_\e\left(4\sqrt{2}\mc{V}^{ij\,M}\bar{\c}_{ijk}\g^\m{\nabla}^-_M\y_\m{}^k\right)\longrightarrow&\ 2\sqrt{2}i\mc{D}_\n\mc{V}_{kl}{}^M\W^{lm}{\nabla}^-_M\bar{\y}_{\m
m}\g^\m\g^\n \e^k,\\
\d_\e\Big(2\sqrt{2}ie\mc{V}_{ij}{}^M\W^{ik}\bar{\y}_{\m
k}\g^{[\m}{\nabla}^+_M(\g^{\n]}\y_\n{}^j)\Big)\longrightarrow& -4\sqrt{2}ie
\W^{ik}\mc{V}_{ij}{}^M{\nabla}_M^+(\bar{\y}_{\m k}\g^{[\m})\g^{\n]}\d_\e
\y_\n^j\\
&-2\sqrt{2}i\, {\nabla}_M^+e\mc{V}_{ij}{}^M\W^{ik}\bar{\y}_{\m
k}\g^{\m\n}\d_\e\y_\n^j\;,\\
\d_\e\Big(\fr{4i}{3}\,{\cal P}_{\m}{}^{ijkl}\bar{\c}_{ijk}\g^\n\g^\m\y_\n^m\W_{lm}\Big)\longrightarrow&\
2\sqrt{2}i \mc{D}_\m\mc{V}_{kl}{}^{M}\W^{lm}\bar{\y}_{\n
m}\g^\m\g^\n {\nabla}^-_M\e^k\\
&-\fr{16}{3}{\cal P}_{\m}{}^{ijkl}\bar{\c}_{ijk}\g^\m{\nabla}^-_M\e^r
\mc{V}_{lr}{}^M{}\\
&-\fr{4\sqrt{2}}{3}{\cal P}_{\m}{}^{ijkl}\bar{\c}_{ijk}\big(\g^\m\g^\n{\nabla}^-
_M\g_\n\big)\e^r\mc{V}_{lr}{}^M{}.
\end{aligned}
\end{equation}
 Let us first separately verify cancellation of the $\F_{\mu\nu}{}^M$ terms  against the variation of the vector kinetic term and of the topological term. 
From the above expressions we have
\begin{equation}
\begin{aligned}
\fr{\sqrt{2}i}{2} e\W^{lm}&\Big( \mc{V}_{kl}{}^M\mc{D}_\m\bar{\y}_{\n
m}\big(\g^{\m\n}\g^{\r\s}-2\g^{\m\n\r}\g^{\s}\big)\e^k\F_{\r\s M}+\fr12\mc{D}_\m \mc{V}_{kl}{}^M\bar{\y}_{\n m}\big(\g^{\r\s}\g^\n\g^\m-\g^\m\g^\n \g^{\r\s}\big)\e^k\F_{\r\s M}\\
&+\mc{V}_{kl}{}^M\bar{\y}_{\n m}\g^{[\m}\g^{\r\s}\g^{\n]}\mc{D}_\m\e^k \F_{M\r\s}\Big)\\
=\fr{\sqrt{2}i}{2} e\W^{lm}&\Big( \mc{V}_{kl}{}^M\mc{D}_\m\bar{\y}_{\n
m}\big(\g^{\m\n\r\s}-2g^{\m\s}g^{\n\r}\big)\e^k\F_{\r\s M}+\mc{D}_\m \mc{V}_{kl}{}^M\bar{\y}_{\n m}\big(-\g^{\m\n\r\s}-2g^{\m\s}g^{\n\r}\big)\e^k\F_{\r\s M}\\
&+\mc{V}_{kl}{}^M\bar{\y}_{\n m}\big(\g^{\m\n\r\s}-2g^{\m\s}g^{\n\r}\big)\mc{D}_\m\e^k \F_{\r\s M}\Big)\\
=\fr{\sqrt{2}i}{2} \W^{lm}&\mc{D}_\m\big( e\mc{V}_{M\,kl}\bar{\y}_{\n
m}\g^{\m\n\r\s}\e^k\big)\F_{\r\s}{}^M+\sqrt{2}i \W^{lm}\mc{D}_\m \Big(\mc{V}_{kl}{}^M\bar{\y}_{\n m}\e^k\Big)\F^{\m\n}{}_{M}
\;,
\end{aligned}
\nonumber
\end{equation}
where we have defined ${\cal F}_{\mu\nu\,M}\equiv {\cal F}_{\mu\nu}{}^N {\cal M}_{MN}$.
The last term above is already present in the $D=5$ reduced theory and cancels the $\bar\e\, \y$ part of the lowest order variation 
of the vector kinetic term. The first term can be rewritten upon partial integration and use of the Bianchi identities (\ref{Bianchi})
\bea
\fr{5\sqrt{2}}{3} \,\W^{lm}\, \varepsilon^{\m\n\r\s\tau} \bar{\y}_{\n
m}\g_\tau \e^k\, \mc{V}_{M\,kl}\,d^{MNK}\partial_N {\cal H}_{\m\r\s\,K}
\;,
\eea
which precisely cancels the corresponding part in the variation (\ref{varBB}) of the topological term. 

To check the remaining terms one first notes the following relations
\begin{equation}
\begin{aligned}
e\g^{\m\n\r}{\nabla}_M\g_\r&=\nabla_M(e\g^{\m\n})\;,\qquad
{\nabla}_M(\g^{\m\n})&=2({\nabla}_M\g^{[\m})\g^{\n]}\;,
\end{aligned}
\end{equation}
which can be used to bring the remainder into the following form
\begin{equation}
\begin{aligned}
&4\sqrt{2}i e \mc{V}_{kl}{}^M\W^{lm}\mc{D}_\m\bar{\y}_{\n m}\g^{\m\n}{{\nabla}}_M\e^k  
+2\sqrt{2}i\mc{V}_{kl}{}^M\W^{lm}\mc{D}_\m\bar{\y}_{\n m}{\nabla}_M(e\g^{\m\n})\e^k\\
&+2\sqrt{2}ie\mc{D}_\m\mc{V}_{kl}{}^M\W^{lm}\Big(-{{\nabla}}_M\bar{\y}_{\n
m}\g^{\m\n}\e^k+ \bar{\y}_{\n m}\g^{\m\n}{{\nabla}}_M\e^k\Big) +2\sqrt{2}ieg^{\m\n}\mc{D}_\m\mc{V}_{kl}{}^M\W^{lm}{\nabla}_M(\bar{\y}_{\n m}\e^k)\\
&-4\sqrt{2}ie \W^{lm}\mc{V}_{kl}{}^M{{\nabla}}_M\bar{\y}_{\n m}\g^{\m\n}\mc{D}_\m\e^k 
-2\sqrt{2}i \mc{V}_{kl}{}^M\W^{lm}\bar{\y}_{\n m}{{\nabla}}_M(e\g^{\m\n})\mc{D}_\m\e^k.
\end{aligned}
\end{equation}
Now integrating by parts of $\mc{D}_\m$ in the first term and of
${\nabla}_M$ in the fourth and the seventh term we get
\begin{equation}
\begin{aligned}
2\sqrt{2}i\W^{lm}&\Big(-2 e \mc{D}_\m\mc{V}_{kl}{}^M\bar{\y}_{\n m}\g^{\m\n}{{\nabla}}_M\e^k 
-2 e \mc{V}_{kl}{}^M\bar{\y}_{\n m}\g^{\m\n}\mc{D}_\m{{\nabla}}_M\e^k 
+\mc{V}_{kl}{}^M\mc{D}_\m\bar{\y}_{\n m}{\nabla}_M(e\g^{\m\n}) \e^k\\
&+e{{\nabla}}_M\mc{D}_\m\mc{V}_{kl}{}^M\bar{\y}_{\n m}\g^{\m\n}\e^k
+\mc{D}_\m\mc{V}_{kl}{}^M\bar{\y}_{\n m}{{\nabla}}_M(e\g^{\m\n})\e^k
+ 2e\mc{D}_\m\mc{V}_{kl}{}^M\bar{\y}_{\n m}\g^{\m\n} {{\nabla}}_M\e^k\\
&+2\mc{V}_{kl}{}^M\bar{\y}_{\n m}{{\nabla}}_M(e\g^{\m\n})\mc{D}_\m\e^k
+2e\mc{V}_{kl}{}^M\bar{\y}_{\n m}\g^{\m\n}{{\nabla}}_M\mc{D}_\m\e^k 
- \mc{V}_{kl}{}^M\bar{\y}_{\n m}{{\nabla}}_M(e\g^{\m\n})\mc{D}_\m\e^k\\
&+eg^{\m\n}\mc{D}_\m\mc{V}_{kl}{}^M{\nabla}_M(\bar{\y}_{\n m}\e^k)\Big)
\;.
\end{aligned}
\end{equation}
Here it is straightforward to construct a commutator from the second terms in the first and the third lines, while the terms with ${\nabla}_M\e^k$ cancel. What is left can be collected into
the following expression
\begin{equation}
\begin{aligned}
2\sqrt{2}i\W^{lm}&\Big(2 e \mc{V}_{kl}{}^M\bar{\y}_{\n m}\g^{\m\n}[{{\nabla}}_M,\mc{D}_\m]\e^k 
+e{{\nabla}}_M\mc{D}_\m\mc{V}_{kl}{}^M\bar{\y}_{\n m}\g^{\m\n}\e^k +eg^{\m\n}\mc{D}_\m\mc{V}_{kl}{}^M{\nabla}_M(\bar{\y}_{\n m}\e^k)\\
&+\mc{V}_{kl}{}^M\mc{D}_\m\bar{\y}_{\n m}{\nabla}_M(e\g^{\m\n}) \e^k 
+\mc{D}_\m\mc{V}_{kl}{}^M\bar{\y}_{\n m}{{\nabla}}_M(e\g^{\m\n})\e^k
+\mc{V}_{kl}{}^M\bar{\y}_{\n m}{{\nabla}}_M(e\g^{\m\n})\mc{D}_\m\e^k\Big)
\;.
\end{aligned}
\nonumber
\end{equation}
Integrating $\mc{D}_\m$ and $\nabla_M$ by parts in the second line this
simplifies into
\begin{equation}
\begin{aligned}
2\sqrt{2}i\W^{lm}&\Big( e \mc{V}_{kl}{}^M\bar{\y}_{\n m}\g^{\m\n}[{{\nabla}}_M,\mc{D}_\m]\e^k 
-e \mc{V}_{kl}{}^M[{{\nabla}}_M,\mc{D}_\m]\bar{\y}_{\n m}\g^{\m\n}\e^k+eg^{\m\n}\mc{D}_\m\mc{V}_{kl}{}^M{\nabla}_M(\bar{\y}_{\n m}\e^k)\Big).
\end{aligned}
\end{equation}
Upon using  the expression \eqref{comm_1} for the commutator of covariant derivatives
together with (\ref{BIR2}), these terms reduce to
\begin{equation}
\begin{aligned}
2\sqrt{2}i\W^{lm}&\Big(-\fr12e \mc{V}_{kl}{}^M\bar{\y}_{\n m}\e^k\widehat{J}^\n{}_M+eg^{\m\n}\mc{D}_\m\mc{V}_{kl}{}^M{\nabla}_M(\bar{\y}_{\n m}\e^k)\Big)\;.
\end{aligned}
\end{equation}
Upon partial integration in the second term, these remaining contributions precisely cancel the corresponding terms in (\ref{varAA}). In what follows we drop the $e$-factor for simpler presentation as the corresponding terms cancel out in the very similar way as above.

\subsection{The \texorpdfstring{$\chi\nabla_M {\cal D}_\mu \e$}{ChiNablaDe} terms}
\label{subsec:DdChi}

There are four fermionic terms from the Lagrangian which contribute such terms
\begin{equation}
\begin{aligned}
&(1)=-\fr{4}{3}\,\bar{\c}_{ijk}\g^\m\mc{D}_\m\c^{ijk}, && (2)=8\sqrt{2}\,i\,\mc{V}_{mn}{}^M\W^{np}\bar{\c}_{pkl} {{\nabla}}^+_M\c^{mkl},\\
&(3)=\fr{4i}{3}\mc{P}_\m{}^{ijkl}\bar{\c}_{ijk}\g^\n\g^\m\y_\n{}^m\W_{lm}, &\quad& (4)=4\sqrt{2}\,\mc{V}_{kl}{}^M\W^{ki}\W^{lj}\bar{\c}_{ijk}\g^\m{\nabla}^-_M\y_\m{}^k.
\end{aligned}
\end{equation}
The relevant terms in supersymmetry variations of these expressions have the following form
\bea
\d^1_\e(1) &=&
-4\sqrt{2}\,\bar{\c}_{ijk}\g^\m\mc{D}_\m ({\cal V}^{ij\,M} {\nabla}_M^- \e^k)\nonumber\\
&=&
-4\sqrt{2}\,\bar{\c}_{ijk}\g^\m {\cal V}^{ij\,M} \mc{D}_\m {\nabla}_M^- \e^k
+4\sqrt{2}\,\bar{\c}_{ijk}\g^\m  {\nabla}_M^- \e^k \,{\cal P}_\mu{}^{ijmn} {\cal V}_{mn}{}^{M}{}\;\nonumber\\
\d^0_\e(2)&=&
-8\sqrt{2}\,\mc{V}_{mn}{}^M\W^{np}\bar{\c}_{pkl}
{\nabla}^+_M ({\cal P}_{\mu}{}^{mklq} \g^\mu \Omega_{qr} \e^r) \nonumber\\
&=&-8\sqrt{2}\,\mc{V}_{mn}{}^M\W^{np}\bar{\c}_{pkl} \g^\mu \Omega_{qr} \e^r\, {\nabla}_M {\cal P}_{\mu}{}^{mklq}
-8\sqrt{2}\,\mc{V}_{mn}{}^M\W^{np}\bar{\c}_{pkl}
 {\cal P}_{\mu}{}^{mklq} \g^\mu \Omega_{qr} {\nabla}_M^+ \e^r\nonumber\\
&&{}-4\sqrt{2}\,\mc{V}_{mn}{}^M\W^{np}\bar{\c}_{pkl}
 \g_\nu \Omega_{qr} \e^r\, {\cal P}_{\mu}{}^{mklq}
 \,{\nabla}_M g^{\mu\nu} 
 +4\sqrt{2}\,\mc{V}_{M\,mn}\W^{np}\bar{\c}_{pkl}
 {\cal P}_{\mu}{}^{mklq} \g_\nu \Omega_{qr} \e^r  
 \F^{\mu\nu}{}^M\;,\nonumber\\
\d^1_\e(3)&=&\ \fr{4\sqrt{2}}{3}\mc{P}_\m{}^{ijkl}\bar{\c}_{ijk}\g^\n\g^\m\,\mc{V}_{rl}{}^{N} 
\left({\nabla}_N^-(\g_\nu\e^r)-\frac13\,\g_\nu {\nabla}_N^- \e^r\right)\nonumber\\
&=&-\fr{4\sqrt{2}}{3}\mc{P}_\m{}^{ijkl}\bar{\c}_{ijk}\g_\nu \e^r\,\nabla_N g^{\mu\nu}  \,\mc{V}_{rl}{}^{N} -\fr{8\sqrt{2}}{3}\,\mc{P}_\m{}^{mijk}{\cal V}{}_{mr}{}^{N} \bar{\c}_{ijk}\, \g^\mu {\nabla}_N^- \e^r\nonumber\\
&&{}
-\fr{2\sqrt{2}}{3}\mc{P}_\m{}^{ijkl}\bar{\c}_{ijk}\,{\cal V}_{M\,rl} 
\F_{\nu\rho}{}^{M}  \g^{\mu\nu\rho} \e^r\;,\nonumber\\
\d^0_\e(4)&=&\ 4\,\sqrt{2}\,\mc{V}^{ij\,M} \bar{\c}_{ijk}\g^\m{\nabla}_M^-{\cal D}_\mu \e{}^k
\;.
\label{var1234}
\eea
The variations of (1) and (4) give rise to a commutator of type \eqref{comm2}
\bea
\lefteqn{
4\sqrt{2}\,\mc{V}^{ij\,M} \bar{\c}_{ijk}\g^\m[{\nabla}_M^-,{\cal D}_\mu] \e{}^k
}
\nonumber\\
\qquad\qquad&=&
8\sqrt{2}\,{\cal V}_{nm}{}^M{}\,\Omega^{mi}  
{\nabla}_M{\cal P}_{\mu}{}^{jknp} \bar{\c}_{ijk}\g^\m \Omega_{pr}  \e^{r}
+2\,{\cal V}_{mn}{}^M{}\,{\nabla}_M{\cal P}^{\mu}{}^{mnij} 
\,{\cal V}_{N\,ij}\, \delta^{(\bar\chi\epsilon)}_\e {\cal A}_\mu{}^N
\qquad\qquad
\nonumber\\
&&{}
+2\sqrt{2}\,{\cal V}_{mn}{}^M{}\,{\cal P}_{\mu}{}^{mnij} 
{\nabla}_M g^{\mu\nu}\,
\bar\chi_{ijk}\g_\nu \e^k
+\sqrt{2}\,\mc{V}^{ij\,M}{\cal R}^-_{M\mu}{}^{ab}\,
\bar{\c}_{ijk}\g^\m \g_{ab}\e^k
\;,
\eea
of which the first term cancels the corresponding term in $\d^0_\e(2)$, and the second term cancels
with the ${\cal J}_M$ contribution in (\ref{varAA}). The $\nabla_M g^{\mu\nu}$ can be seen to cancel
against the contributions from $\d^0_\e(2)$ and $\d^1_\e(3)$ by virtue of the ${\rm USp}(8)$ identity (\ref{ident3}).
By the same identity, the three $\nabla_M\epsilon$ terms in (\ref{var1234}) would cancel if they came with the same
spin connection $\omega_M^-$, i.e.\ they induce an extra term in the field strength ${\cal F}_{\mu\nu}{}^M$.

Collecting all resulting terms, we arrive at
\bea
\mbox{(\ref{var1234})} &=&
\sqrt{2}\,\mc{V}^{ij\,M}{\cal R}^-_{M\mu}{}^{ab}\,\bar{\c}_{ijk}\g^\m \g_{ab}\e^k
-2\sqrt{2}\,\mc{V}_{M\,mn}\W^{np}\bar{\c}_{pkl}
 {\cal P}_{\mu}{}^{mklq} \g^\mu \Omega_{qr} \gamma^{\nu\rho}   \e^r\,{\cal F}_{\nu\rho}{}^M
\label{A10}\\
&&{}+4\sqrt{2}\,\mc{V}_{M\,mn}\W^{np}\bar{\c}_{pkl}
 {\cal P}_{\mu}{}^{mklq} \g_\nu \Omega_{qr} \e^r  
 \F^{\mu\nu}{}^M
 -\fr{2\sqrt{2}}{3}\mc{P}_\m{}^{ijkl}\bar{\c}_{ijk}\,{\cal V}_{M\,rl} 
\F_{\nu\rho}{}^{M}  \g^{\mu\nu\rho} \e^r
\;, \nonumber
\eea
with the second term coming from converting $\nabla_M^+$ into $\nabla_M^-$.
Now the curvature term can be expanded with (\ref{BianchiMix}), (\ref{BIR2}) as
\bea
\sqrt{2}\,\mc{V}^{ij\,M}{\cal R}^-_{M\mu}{}^{ab}\,\bar{\c}_{ijk}\g^\m \g_{ab}\e^k&=& 
\sqrt{2}\,\mc{V}^{ij\,M}{\cal R}^-_{M\mu\nu\rho}\,\bar{\c}_{ijk}\g^{\mu\nu\rho}\e^k+2\sqrt{2}\,\mc{V}^{M \,ij}{\cal R}^-_{M\mu}{}^{\mu\nu}\,
\bar{\c}_{ijk}\g_\nu\e^k\nonumber\\
&=& 
\fr{1}{\sqrt{2}}\,{\cal D}_{[\mu}(\F_{\nu\rho]}{}^N{\cal M}_{NM})
\,\mc{V}^{ij\,M} \bar{\c}_{ijk}\g^{\mu\nu\rho}\e^k\\
&&{}+\sqrt{2}\,\mc{V}^{ij\,M}\widehat{J}^\mu{}_M \,
\bar{\c}_{ijk}\g_\mu\e^k
-\sqrt{2}\,\mc{V}^{ij\,M}
e_a{}^\nu e_b{}^\mu\, {\cal D}_\mu({\cal M}_{MN}\F^{ab\,N})
\bar{\c}_{ijk}\g_\nu\e^k\;. \nonumber
\eea
The last two terms cancel against the vector field variation from the Einstein-Hilbert term (\ref{varAA}) and from the vector kinetic term. 
The first term gives
\begin{equation}\begin{aligned}
\label{ll3}
\rightarrow& \
\frac{1}{\sqrt{2}}\,\mc{V}_M{}^{ij}\,{\cal D}_{[\mu}\F_{\nu\rho]}{}^M
\, \bar{\c}_{ijk}\g^{\mu\nu\rho}\e^k
+
\sqrt{2}\,{\cal P}_{\mu}{}^{ijmn} {\cal V}_{M\,mn}\F_{\nu\rho}{}^M\, 
\bar{\c}_{ijk}\g^{\mu\nu\rho}\e^k
\\
=&\ \frac{1}{\sqrt{2}}\,\mc{V}_M{}^{ij}\,{\cal D}_{[\mu}\F_{\nu\rho]}{}^M
\, \bar{\c}_{ijk}\g^{\mu\nu\rho}\e^k
\\
&{}
+
\frac23\sqrt{2}\, {\cal V}_M{}_{mn}  {\cal P}_{\mu}{}^{mijk} \F_{\nu\rho}{}^M
 \bar{\c}_{ijk}\g^{\mu\nu\rho} \e^n
-2\sqrt{2}\, {\cal V}_M{}_{mn} \F_{\nu\rho}{}^M \Omega_{pr} 
\bar{\c}_{ijk} \g^{\mu\nu\rho}\e^r \Omega^{mi}  {\cal P}_{\mu}{}^{jknp}
\end{aligned}\end{equation}
where we have once more used the algebraic identity (\ref{ident3}).
The last two terms precisely cancel the ${\cal F}_{\mu\nu}{}^M$ terms from (\ref{A10}).
We remain with the first term of (\ref{ll3}) which can be rewritten with the Bianchi identity (\ref{Bianchi})
and cancels against the corresponding ${\cal H}_{\mu\nu\rho\,M}$ term from variation of the topological term in (\ref{varBB}).

\subsection{The \texorpdfstring{$\psi\nabla_M\nabla_N\e$}{PsiNablaNabla e} terms}
\label{subsec:DDPsi}

These terms arise from the $\nabla_M\epsilon$ variation of the following two terms from
the Lagrangian (\ref{L}) 
\begin{equation}
\begin{aligned}
(1)=2\sqrt{2}\,i\,\mc{V}_{ij}{}^M\W^{ik}\bar{\y}_{\m k}\g^{[\m}{\nabla}_M^+\left(\g^{\n]}\y_\n{}^j\right), && 
(2)=4\sqrt{2}\,\mc{V}_{ij}{}^M\bar\y_\m{}_k{\nabla}^-_M \left( \g^\m {\c}^{ijk} \right).
\end{aligned}
\end{equation}
Explicitly, with \eqref{transf} this gives
\begin{equation}\begin{aligned}
\delta^1_\e (1) =&\
8\,\mc{V}_{nj}{}^M \mc{V}^{jk\,N} 
\bar{\y}_{\m k}\g^{[\m}{\nabla}_N^+\left(\g^{\n]}
\left({\nabla}_M^-(\g_\n\e^n)-\fr13\g_\n{\nabla}_M^-\e^n\right)\right)
\;,
\\
\delta^1_\e(2) =&
-12\,\left(\mc{V}^{[ij\,N}\W^{k]m}-\fr13\mc{V}^{m[i\,N}\W^{jk]}\right)\W_{mr}
\mc{V}_{ij}{}^M\bar\y_\m{}_k{\nabla}_M^- \left( \g^\m 
{\nabla}_N^-\e^r
 \right)
\\
=&\
4\,\left({\cal M}^{MN}\,\delta^k_n+2\mc{V}_{nj}{}^M\mc{V}^{jk\,N} 
+\fr23\mc{V}_{ij}{}^M\mc{V}^{jm\,N}\W^{ik}\W_{mn} \right)
\bar\y_\m{}_k{\nabla}_M^- \left( \g^\m 
{\nabla}_N^-\e^n
 \right)
 \\
=&\
4\,\left({\cal M}^{MN}\,\delta^k_n+2\mc{V}_{nj}{}^M\mc{V}^{jk\,N} 
+\fr23\mc{V}_{nj}{}^N \mc{V}^{jk\,M} \right)
\bar\y_\m{}_k{\nabla}_M^- \left( \g^\m 
{\nabla}_N^-\e^n
 \right)
 \;.
\end{aligned}\end{equation}
Let us now consider the terms containing $\nabla_M$ and the gauge field flux $\F_{\m\n}{}^M$ separately. For the derivative terms 
and ignoring all derivatives on the external metric we have
\begin{equation}\begin{aligned}
\delta_\e (1)+\delta_\e (2) \rightarrow&
\left(\frac{64}{3}+\frac83\right)\,\mc{V}_{nj}{}^M \mc{V}^N{}^{jk} 
\bar{\y}_{\m k}\g^{\m}{\nabla}_N {\nabla}_M\e^n
\\
&{}
+8\,\mc{V}_{nj}{}^M\mc{V}^{jk\,N}\bar{\y}_{\m k}\g^{\m}{\nabla}_M {\nabla}_N\e^n
+4 {\cal M}^{MN}\,\bar{\y}_{\m k}\g^{\m}{\nabla}_M {\nabla}_N\e^k
\\
=&
\left(-\frac{32}{3}-\frac43+4\right)\,\mc{V}_{nj}{}^M \mc{V}{}^{jk\,N} 
\bar{\y}_{\m k}\g^{\m}[{\nabla}_M, {\nabla}_N]\e^n
\\
&{}
\left(+\frac{64}{3}+\frac83+8\right)\,\mc{V}_{nj}{}^M\mc{V}^{jk\,N}
\bar{\y}_{\m k}\g^{\m}{\nabla}_{(M} {\nabla}_{N)}\e^n
+4 {\cal M}^{MN}\,\bar{\y}_{\m k}\g^{\m}{\nabla}_M{\nabla}_N\e^k
\\
=&
-8\,\mc{V}_{nj}{}^M \mc{V}^{jk\,N} 
\bar{\y}_{\m k}\g^{\m}[{\nabla}_M, {\nabla}_N]\e^n
\\
&{}
+32\,\mc{V}_{nj}{}^M\mc{V}^{jk\,N}
\bar{\y}_{\m k}\g^{\m}{\nabla}_{(M} {\nabla}_{N)}\e^n
+4 {\cal M}^{MN}\,\bar{\y}_{\m k}\g^{\m}{\nabla}_M {\nabla}_N\e^k
\\
=&
-\frac12\,{\cal R}\,\bar\psi_{\mu\,k} \g^\mu \e^k 
\;,
\end{aligned}\end{equation}
upon using (\ref{comm_nabla_1}).

That cancels the corresponding variations of the scalar potential. Now for the $\F\F$ terms altogether we obtain
\begin{equation}\begin{aligned}
\d_\e(1)+\d_\e(2)\rightarrow&\
  \frac1{16}\left({\cal M}^{MN}\,\delta^k_n
+8\,\mc{V}_{nj}{}^N{}\mc{V}^{jk\,M} \right)
\bar{\y}_{\m k}\g^{\mu\kappa\lambda\rho\sigma}\e^n\,\F_{\kappa\lambda\,M} \F_{\rho\sigma\,N}
\\
&{}
-\frac18\,{\cal M}^{MN}\,
\bar{\y}_{\m k}\g^{\mu}\e^k\,\F_{\rho\nu\,M} \F^{\rho\nu}{}_{N}
-\frac12\,{\cal M}^{MN}\,
\bar{\y}_{\m k}\g^{\rho}\e^k\,\F^{\mu\nu}{}_{M} \F_{\nu\rho}{}_{N}
\\
=&
 - \frac{i}{4}\sqrt{5}\,\varepsilon^{\mu\kappa\lambda\rho\sigma} d^{MNK}\,
  \, {\cal V}_{ij}{}^K{} \Omega^{ir} \bar\e_r \psi_\mu{}^j \,\F_{\kappa\lambda}{}^M \F_{\rho\sigma}{}^{N}
\\
&{}
+\frac14\,(e^{-1} \delta_\e e)\, {\cal M}^{MN}\,\F_{\rho\nu\,M} \F^{\rho\nu}{}_{N}
+\frac14\,\delta_\e(\g^{\mu\nu}g^{\rho\sigma})\,{\cal M}^{MN}\,
\F_{\mu\rho}{}_{M} \F_{\nu\sigma}{}_{N},
\end{aligned}\end{equation}
that precisely cancels the variation from the kinetic and topological vector term. 
Let us turn to the terms of the form $\nabla \F$ that give
\begin{equation}\begin{aligned}
 \d_\e(1)+\d_\e(2)\rightarrow&\
 \mc{V}_{nj}{}^M \mc{V}^{jk\,N} 
\bar{\y}_{\m k} \left(
\g^{\rho\sigma}\g^\mu\e^n
+\fr13\g^{\mu}\g^{\rho\sigma} \e^n\right) \nabla_N \F_{\rho\sigma\,M}
\\
&{}
-\frac12\left({\cal M}^{MN}\,\delta^k_n+2\mc{V}^N{}_{nj}\mc{V}^{jk\,M} 
+\fr23\mc{V}_{nj}{}^M\mc{V}^{jk\,N} \right)
\bar\y_\m{}_k \g^\m 
\g^{\rho\sigma} \e^n
   \nabla_N \F_{\rho\sigma\,M}
\\
=&
-\frac12\,{\cal M}^{MN}\,\bar\y_\m{}_k 
\g^{\mu\rho\sigma} \e^k
   \nabla_M \F_{\rho\sigma\,N}
  -2\,\mc{V}_{nj}{}^M\mc{V}^{jk\,N} \bar\y_\m{}_k 
\g^{\mu\rho\sigma} \e^n   \nabla_{[M} \F_{\rho\sigma\,N]}
\\
&{}
-4\,\mc{V}_{nj}{}^M\mc{V}^{jk\,N} \,\bar\y_\mu{}_k \g_{\nu} \e^n
   \nabla_{(M} \F^{\mu\nu}{}_{N)}
-\frac12\,{\cal M}^{MN}\,\,\bar\y_\mu{}_k \g_{\nu} \e^k
   \nabla_{M} \F^{\mu\nu}{}_{N}
\\
=&
  -2\,\mc{V}_{nj}{}^M\mc{V}^{jk\,N} \bar\y_\m{}_k 
\g^{\mu\rho\sigma} \e^n   \nabla_{[M} \F_{\rho\sigma\,N]}
-\frac12\,\bar\y_\m{}_k 
\g^{\mu\rho\sigma} \e^k
   \nabla_M \F_{\rho\sigma}{}^M
\\
&{}
-5\,d^{MNK} \,\delta^{(\bar\psi\epsilon)}_\e {\cal B}_{\mu\nu\,K}\, 
\nabla_{(M} \F^{\mu\nu}{}_{N)}\;.
\end{aligned}\end{equation}
The first line here precisely cancels against the corresponding 
terms in (\ref{varRS}) from variation of the Rarita-Schwinger term.
The second line cancels against the corresponding contribution in (\ref{varBB}).
Finally, for the terms of type $\F\nabla\e$, we obtain
\begin{equation}\begin{aligned}
\delta_\e (1) \rightarrow&\
\mc{V}_{nj}{}^M \mc{V}^{jk\,N} 
\bar{\y}_{\m k}\g^{[\m}\g^{\rho\sigma}\left(\g^{\n]}
\left(\fr23\g_\n{\nabla}_M\e^n\right)\right) \F_{\rho\sigma\,N}
\\
&{}
 -\mc{V}_{nj}{}^N{} \mc{V}^{jk\,M} 
\bar{\y}_{\m k}\g^{[\m}\left(\g^{\n]}
\left(\g^{\rho\sigma} \g_\n-\fr13\g_\n\g^{\rho\sigma}\right)\right){\nabla}_M\e^n  \F_{\rho\sigma\,N}
\\
=&\
\fr43\mc{V}_{nj}{}^M \mc{V}^{jk\,N} 
\bar{\y}_{\m k}\g^{\m\rho\sigma}{\nabla}_M\e^n \F_{\rho\sigma\,N}
+
\fr{4}3\mc{V}_{nj}{}^M \mc{V}{}^{jk\,N} 
\bar{\y}_{\m k}\g^{\sigma}{\nabla}_M\e^n \F^\mu{}_{\sigma\,N}
\\
&{}
 +\frac43\mc{V}_{nj}{}^N{} \mc{V}^{jk\,M} 
\bar{\y}_{\m k}
\g^{\m\rho\sigma} {\nabla}_M\e^n  \F_{\rho\sigma\,N}
-\frac43 \mc{V}_{nj}{}^N{} \mc{V}^{jk\,M} 
\bar{\y}_{\m k}
\g^{\sigma}{\nabla}_M\e^n  \F^\mu{}_{\sigma\,N}\;,\\
\delta_\e(2)\rightarrow&
-\frac12\left({\cal M}^{MN}\,\delta^k_n+2\mc{V}_{nj}{}^N{}\mc{V}^{jk\,M} 
+\fr23\mc{V}_{nj}{}^M\mc{V}^{N jk} \right)
\bar\y_\m{}_k\g^{\rho\sigma} \g^\m {\nabla}_M\e^n \F_{\rho\sigma\,N}
\\
&{}
-\frac12\,\left({\cal M}^{MN}\,\delta^k_n+2\mc{V}_{nj}{}^M\mc{V}^{jk\,N} 
+\fr23\mc{V}_{nj}{}^N \mc{V}^{jk\,M} \right)
\bar\y_\m{}_k \g^\m 
\g^{\rho\sigma} {\nabla}_M \e^n
 \F_{\rho\sigma\,N}.
\end{aligned}\end{equation}
Together these contribution simplify to the following nice expression
\begin{equation}\begin{aligned}
\delta_\e(1) + \delta_\e(2) &\rightarrow
-\,\bar\y_\m{}_k\g^{\mu \rho\sigma} \F_{\rho\sigma}{}^M  {\nabla}_M\e^k,
\end{aligned}\end{equation}
which precisely cancels the corresponding contribution from (\ref{varRS}).

\subsection{The \texorpdfstring{$\chi\nabla_M\nabla_N\e$}{ChiNablaNabla e} terms}
\label{subsec:DDChi}

As the final check we collect the $\chi\nabla_M\nabla_N\e$ terms 
which originate from the $\nabla_M\epsilon$ variation
of the following two terms
\begin{equation}
\begin{aligned}
(1)=8\sqrt{2}\,i\,\mc{V}_{mn}{}^M\W^{np}\bar{\c}_{pkl}{\nabla}_M^+ \c^{mkl}, && 
(2)=4\sqrt{2}\,\mc{V}_{ij}{}^M\bar\y_\m{}_k{\nabla}_M^- \left( \g^\m {\c}^{ijk} \right)\;,
\end{aligned}
\end{equation}
of the Lagrangian (\ref{L}).
Their supersymmetry variation gives 
\begin{equation}\begin{aligned}
\label{l2}
\delta_\e(1)=&
-48\,i\,\mc{V}_{mn}{}^M\W^{np}
\left(\mc{V}^{N[lm}\W^{k]j}-\fr13\mc{V}^{N j[l}\W^{mk]}\right)\W_{jr}
\bar{\c}_{pkl}{\nabla}_M^+{\nabla}_N^-\e^r
\\
=&
16\,i 
\left(2\mc{V}^{kj\,N} \Omega_{jn} \mc{V}{}^{nl\,M} \delta^{p}_r
+\mc{V}^{kl\,N} \mc{V}^{pj\,M} \Omega_{jr} 
+\fr23 \mc{V}{}^{kl\,M}\mc{V}^{pj\,N} \W_{jr}  \right)
\bar{\c}_{pkl}{\nabla}_M^+{\nabla}_N^-\e^r,\\
\delta_\e(2)=&
-8i\,\mc{V}{}^{kl\,M} \mc{V}^{pj\,N} \W_{jr}\,
\bar{\c}_{klp}\g^\m{\nabla}_M^-
\left(\fr23\g_\m{\nabla}_N^-\e^r
+ \g_a \e^r \,{\nabla}_N^- e_\mu{}^a\right)
\;.
\end{aligned}\end{equation}
Again for simplicity we start from analysis for the terms that do not contain the field strength 
\begin{equation}\begin{aligned}
\label{ll4}
\delta_\e (1)+\delta_\e (2) \rightarrow&\
32\,i \,\mc{V}^{kj\,M} \Omega_{jn} \mc{V}^{nl\,N} \,
\bar{\c}_{pkl}\nabla_{(M}\nabla_{N)}\e^p+
16\,i 
\mc{V}^{kl\,N} \mc{V}^{pj\,M} \Omega_{jr} 
\bar{\c}_{pkl}[\nabla_M,\nabla_N]\e^r
\\ 
&-8i\,\mc{V}{}^{kl\,M} \mc{V}^{pj\,N} \W_{jr}\,
\bar{\c}_{klp}\g^\m
 \g_a \e^r \,{\nabla}_M {\nabla}_N e_\mu{}^a\\
=& \
16\,i \left(
\mc{V}^{kl\,N} \mc{V}{}^{pj\,M} \Omega_{jr} 
\bar{\c}_{pkl}[\nabla_M,\nabla_N]\e^r +
2\,\mc{V}^{kj\,M} \Omega_{jn} \mc{V}{}^{nl\,N} \,
\bar{\c}_{pkl}\nabla_{(M}\nabla_{N)}\e^p
\right)
\\
\stackrel{(\ref{comm2})}{=}&
-4i\,{\cal R}^{ijkl}\, \Omega_{ri} 
\bar{\c}_{jkl} \e^r
+2i\,\mc{V}^{kl\,M} \mc{V}^{pj\,N} \W_{jr}\,
\bar{\c}_{klp} \e^r \nabla_M g_{\mu\nu} \nabla_N g^{\mu\nu} 
\;,
\end{aligned}\end{equation}
with the curvature ${\cal R}^{ijkl}$ in the ${\bf 42}$ representation. These terms,
after using the section constraint for the second one, precisely cancel 
the variation of the scalar potential (\ref{varV}). 
Collecting now the $\nabla\F$ terms, we get
\begin{equation}\begin{aligned}
\label{nabF}
\delta_\e (1)+\delta_\e (2) \rightarrow&
- 4i\,\mc{V}{}^{kl\,M} \mc{V}_K{}^{pj} \W_{r[k}\,
\bar{\c}_{lpj]}\g^{\mu\nu} \e^r \, \nabla_{M}  \F_{\mu\nu}{}^K
- 4i\,\mc{V}{}^{kl\,M} \mc{V}_K{}^{pj} \W_{r[k}\,
\bar{\c}_{lpj]}\g^\mu{}_{\lambda} \e^r \,  \F^{\lambda\rho}{}^K\,\nabla_{M} g_{\mu\rho} 
\\
&{}
-4i\,\mc{V}^{kj\,M} \Omega_{jn} \mc{V}_K{}^{nl} \,
\bar{\c}_{pkl} \g^{ab} \e^p\,{\nabla}_{M} \F_{ab}{}^K
\\
=&
- 4i\,\mc{V}{}^{kl\,M} \mc{V}_K{}^{pj} \W_{r\llbracket k}\,
\bar{\c}_{lpj\rrbracket}\g^{\mu\nu} \e^r \, \nabla_{M}  \F_{\mu\nu}{}^K
- 4i\,\mc{V}{}^{kl\,M} \mc{V}_K{}^{pj} \W_{r[k}\,
\bar{\c}_{lpj]}\g^\mu{}_{\lambda} \e^r \,  \F^{\lambda\rho}{}^K\,\nabla_{M} g_{\mu\rho} 
\\
&{}
-2i\,\mc{V}^{kj\,M} \Omega_{jn} \mc{V}_K{}^{nl} \,
\bar{\c}_{pkl} \g_{\mu\nu} \e^p\,{\nabla}_{M} \F^{\mu\nu}{}^K
\\
=&
- 4i\,\mc{V}{}^{kl\,M} \mc{V}_K{}^{pj} \W_{r\llbracket k}\,
\bar{\c}_{lpj\rrbracket}\g^{\mu\nu} \e^r \, \nabla_{M}  \F_{\mu\nu}{}^K
- 4i\,\mc{V}{}^{kl\,M} \mc{V}_K{}^{pj} \W_{r[k}\,
\bar{\c}_{lpj]}\g^\mu{}_{\lambda} \e^r \,  \F^{\lambda\rho}{}^K\,\nabla_{M} g_{\mu\rho} 
\\
&{}
-5\,d^{KMN}\,\delta_\e B_{\mu\nu\,N}
\,{\nabla}_{M}({\cal M}_{KL} \F^{\mu\nu}{}^L)\;.
\end{aligned}\end{equation}
The last term precisely cancels the corresponding variation of the vector kinetic term, the second term cancels against (\ref{lastT}) below, the first one
upon using the identity (\ref{id7}) cancels against the contribution from (\ref{varF}).
Collecting the $\F\F$ terms we obtain (again with ${\cal F}_M\equiv {\cal M}_{MN}{\cal F}^N$)
\begin{equation}\begin{aligned}
\delta_\e (1)+\delta_\e (2) \rightarrow&
-\frac14\,i 
\left(2\mc{V}^{kj\,N} \Omega_{jn} \mc{V}{}^{nl\,M} \delta^{p}_r
+\mc{V}^{kl\,N} \mc{V}{}^{pj\,M} \Omega_{jr} 
+\fr23 \mc{V}{}^{kl\,M}\mc{V}^{pj\,N} \W_{jr}  \right)
\bar{\c}_{pkl}\g^{\mu\nu}\g^{\rho\sigma}\e^r\,\F_{\mu\nu\,M} \F_{\rho\sigma\,N}
\\
&{}
-\frac18\,i\,\mc{V}{}^{kl\,M} \mc{V}^{pj\,N} \W_{jr}\,
\bar{\c}_{klp}
\left(\g^\tau \g^{\mu\nu}\g^{\rho\sigma} \g_\tau -\fr13 \g^\tau \g^{\mu\nu} \g_\tau \g^{\rho\sigma}\right)
 \e^r
\\
=&
 -\frac{i}4 
\left(2\mc{V}^{kj\,N} \Omega_{jn} \mc{V}{}^{nl\,M} \delta^{p}_r
+\mc{V}^{kl\,N} \mc{V}{}^{pj\,M} \Omega_{jr} 
 \right)
\bar{\c}_{pkl}(\g^{\mu\nu\rho\sigma}+4\g^{\mu\sigma} g^{\nu\rho}-2g^{\mu\rho}g^{\nu\sigma})\e^r\,\F_{\mu\nu\,M} \F_{\rho\sigma\,N}
\\
&{}
-\frac{i}8\,\mc{V}{}^{kl\,M} \mc{V}^{pj\,N} \W_{jr}\,
\bar{\c}_{klp}
(-2\, \g^{\mu\nu\rho\sigma}+8\,\g^{\mu\sigma} g^{\nu\rho}-12\,g^{\mu\rho}g^{\nu\sigma})
 \e^r\,\F_{\mu\nu\,M} \F_{\rho\sigma\,N}
\\
=&
 -\frac{i}2\,\mc{V}^{kj\,N} \Omega_{jn} \mc{V}{}^{nl\,M} \delta^{p}_r
\,
\bar{\c}_{pkl} \g^{\mu\nu\rho\sigma} \e^r\,\F_{\mu\nu}{}^M \F_{\rho\sigma}{}^{N}
+2i\,\mc{V}_M{}^{kl} \mc{V}_N{}^{pj} \W_{r\llbracket k}\,
\bar{\c}_{lpj\rrbracket}\, \e^r\,\F_{\mu\nu}{}^M \F^{\mu\nu\,N}.
\end{aligned}\end{equation}
These cancel against the corresponding variation of the kinetic and the topological vector terms.
Finally, for the $\F\nabla$ terms, we write
\begin{equation}\begin{aligned}
\delta_\e (1)+\delta_\e (2) \rightarrow&\
\frac23i 
\mc{V}{}^{kl\,N}\mc{V}_M{}^{pj} \W_{jr} \F_{\mu\nu}{}^M
\bar{\c}_{pkl}\g^{\mu\nu}{\nabla}_N\e^r
-\frac23i 
\mc{V}_{M}{}^{kl}\mc{V}^{pj\,N} \W_{jr}  \F_{\mu\nu}{}^M
\bar{\c}_{pkl}\g^{\mu\nu}{\nabla}_N\e^r
\\
&{}
+i\,\mc{V}{}^{kl\,M} \mc{V}^{pj\,N} \W_{jr}\,
\bar{\c}_{klp}\g^{\nu\rho}
{\nabla}_N \e^r\F_{\nu\rho\,M}
-\frac{i}3\,\mc{V}{}^{kl\,M} \mc{V}^{pj\,N} \W_{jr}\,
\bar{\c}_{klp}\g^{\nu\rho}{\nabla}_N\e^r \F_{\nu\rho\,M}
\\
&{}
+i\,\mc{V}{}^{kl\,N} \mc{V}^{pj\,M} \W_{jr}\,
\bar{\c}_{klp}\
 \g^{\nu\rho} {\nabla}_N \e^r \F_{\nu\rho\,M}
-\frac{5i}3\,\mc{V}{}^{kl\,N} \mc{V}^{pj\,M} \W_{jr}\,
\bar{\c}_{klp}\g^{\nu\rho} {\nabla}_N \e^r \F_{\nu\rho\,M}
\\
&{}
+i\,\mc{V}{}^{kl\,M} \mc{V}^{pj\,N} \W_{jr}\,
\bar{\c}_{klp}\g^\m\g^{\nu\rho}
\g_a \e^r\F_{\nu\rho\,M}\,\nabla_N e_\mu{}^a
\\
&{}
-\frac{i}3\,\mc{V}{}^{kl\,N} \mc{V}^{pj\,M} \W_{jr}\,
\bar{\c}_{klp}\g^\m\g_a \g^{\nu\rho} \e^r \F_{\nu\rho\,M} \nabla_N e_\mu{}^a
\\
=&
4i\,\mc{V}{}^{kl\,M} \mc{V}^{pj\,N} \W_{jr}\,
\bar{\c}_{klp}\g^{\mu\nu} \e^r\F_{\nu\rho\,(M}\,g^{\rho\lambda} \nabla_{N)} g_{\mu\lambda}
\;,
\label{lastT}
\end{aligned}\end{equation}
that precisely cancels the second term above in (\ref{nabF}).

\section{${\rm \mathbf{USp(8)}}$ Identities}
\label{app:identities}

In this section some useful algebraic relations, that follow from the structure of ${\rm USp}(8)$ representations. 
Their derivation was facilitated in part by using the computer algebra system {\tt Cadabra} \cite{Peeters:2006kp,Peeters:2007wn}. 
Some of the more complicated algebraic relations were obtained using an explicitly chosen ${\rm USp}(8)$ representation.

We first recall the notation of double brackets
\bea
P^{\llbracket ijkl \rrbracket} &=& P^{[ijkl]} - (\Omega\mbox{-traces})\;,\qquad \mbox{etc.}\;,
\eea
in order to define the irreducible ${\rm USp}(8)$ representations. 
E.g.\ the tensor $P^{ijkl}=P^{\llbracket ijkl\rrbracket}$ defines the irreducible ${\bf 42}$ 
representation of ${\rm USp}(8)$ and can explicitly be constructed by making use of
the corresponding projector
\bea
P^{\llbracket ijkl\rrbracket} &=& \mathbb{P}_{{\bf 42}}^{ijkl}{}_{mnpq}\,P^{mnpq} \;,\nonumber\\
&&{}
\mathbb{P}_{{\bf 42}}^{ijkl}{}_{mnpq} ~\equiv~
\delta^{ijkl}{}_{mnpq} - \frac32\,\Omega^{[ij}\delta^{kl]}{}_{[mn}\Omega_{pq]}
+\frac18\,\Omega^{[ij}\Omega^{kl]}\Omega_{[mn}\Omega_{pq]}
\;.
\label{P42}
\eea

Several of the ${\rm USp}(8)$ identities are not straightforward to derive but most conveniently derived by
identifying the underlying representation structure. 
A simple example of such an identity is
\bea
P^{[ijkl} \Omega^{mn]}&=& 0\;,
\label{idi1}
\eea
for $P^{ijkl}=P^{\llbracket ijkl\rrbracket}$ in the ${\bf 42}$ of ${\rm USp}(8)$.
The identity (\ref{idi1}) follows straightforwardly from the fact that there is no ${\bf 42}$
representation in the six-fold antisymmetric tensor product.

In the same manner, one may derive the identity 
\bea
0&=&
\frac34\,{\cal V}{}_{mn}{}^M \epsilon^{\llbracket i} P^{jk\rrbracket mn}
-\frac12\, {\cal V}_{mn}{}^M  P^{mijk} \epsilon^n
+\frac32\, {\cal V}_{mn}{}^M \Omega_{pr}\epsilon^r \Omega^{m\llbracket i}  P^{jk\rrbracket np}
\;,
\label{ident3}
\eea
for $P^{ijkl}=P^{\llbracket ijkl\rrbracket}$,
whose existence follows from the fact that there is no ${\bf 42}$ representation in the tensor product ${\bf 27}\times{\bf 42}$,
and as a consequence there are only two singlets in ${\bf 8}\otimes {\bf 27}\otimes {\bf 48} \otimes {\bf 42}$\,.
The coefficients in (\ref{ident3}) can then be fixed by employing an explicit realization of these objects,
or by using the explicit form (\ref{P42}) of the projector.

Similarly, one shows the identity
\begin{equation}\begin{aligned}
 {\cal V}_K{}^{\llbracket ij}\,{\cal V}^{kl\rrbracket M}\,
\Omega_{ri} \bar\chi_{jkl}
=&\
{\cal V}^{mn\,M} {\cal V}_K{}^{pq}\,\left(\Omega_{r[m} \bar\chi_{npq]}
 - \frac32\, \Omega_{ri} \bar\chi_{jkl}\,\Omega^{[ij}\delta^{kl]}{}_{[mn}\Omega_{pq]}\right)
\\
=&\
{\cal V}^{mn\,M} {\cal V}_K{}^{pq}\,\Omega_{r[m} \bar\chi_{npq]}
 + \frac12\, {\cal V}^{mn\,M} \Omega_{nq} {\cal V}_K{}^{qp}\,\bar\chi_{rmp}\;.
 \label{id7}
\end{aligned}\end{equation}

In the main text, the calculation of the scalar potential 
(\ref{Vcov}) and its properties like (\ref{varR1}) require
further ${\rm USp}(8)$ identities.
E.g.\ one derives
\begin{equation}\begin{aligned}
 {\cal V}_{ij}{}^{M} {\cal V}_{kl}{}^{N} \partial_{(M} p_{N)}{}^{jklm}\, \e_m-
  \Omega_{im}\,{\cal V}_{pj}{}^{M} {\cal V}_{kl}{}^{N} \partial_{(M} p_{N)}{}^{jklm}\, \Omega^{pq} \e_{q}
=& -\frac14\,  {\cal V}_{kl}{}^{M} {\cal V}_{mn}{}^{N} \partial_{(M} p_{N)}{}^{klmn}\, \e_i
\\
&{}
+\Omega_{ir} {\cal V}_{km}{}^{M}\Omega^{mn}{\cal V}_{nl}{}^N\,
\partial_{(M} p_{N)}{}^{rjkl}\, \e_j
\;,
\end{aligned}\end{equation}
that follows from the fact that in the above
${\cal V}_{ij}{}^{M} {\cal V}_{kl}{}^{N}$ appears only projected onto the $\bf 42$ due to the section condition (\ref{section2}),
and furthermore there is only $\bf 1$ and no $\bf 36$ in $\bf (42\otimes42)_{\rm sym}$.

Another set of relations is required for the evaluation of the commutator (\ref{comm_nabla_2}) that contains
\begin{equation}\begin{aligned}
\rightarrow&\ 
\mc{V}_{mn}{}^M {\cal V}^{kl\,N} {\Omega}_{p q} {\e}^{p} {\chi}_{k l r} {p}_{M N}\,^{m n q r} 
- {\cal V}^{ij\,M} {\cal V}^{kl\,N} {\Omega}^{m n} {\e}^{p} {\chi}_{i k m} {p}_{M N\, j l n p} 
\\
&{}
- \frac{1}{2}\, \mc{V}_{mn}{}^M {\cal V}^{kl\,N} {\Omega}_{k p} {\e}^{q} {\chi}_{l q r} {p}_{M N}\,^{m n p r} 
- \frac{1}{16}\, {\cal M}^{M N} {\Omega}_{i j} {\e}^{i} {\chi}_{k l m} {p}_{M N}\,^{j k l m} 
\\
&{}
+ \frac{1}{4}\, \mc{V}_{mn}{}^M {\cal V}^{kl\,N} {\Omega}_{k p} {\e}^{p} {\chi}_{l q r} {p}_{M N}\,^{m n q r} 
+ \frac{3}{2}\, \mc{V}_{mn}{}^M {\cal V}^{kl\,N} {\Omega}_{k p} {\e}^{m} {\chi}_{l q r} {p}_{M N}\,^{n p q r}
\;,
\label{rescomm}
\end{aligned}\end{equation}
where we denote $p_{MN}{}^{ijkl}\equiv \partial_{(M} p_{N)}{}^{ijkl}$\,.
Next, one notes a non-trivial ${\rm USp}(8)$ identity
\begin{equation}\begin{aligned}
0=&\
\frac12\,\mc{V}_{mn}{}^M {\cal V}^{k l\,N} {\Omega}_{p q} {\e}^{p} {\chi}_{k l r} {p}_{M N}\,^{m n q r} 
- {\cal V}^{M i j} {\cal V}^{k l\,N} {\Omega}^{m n} {\e}^{p} {\chi}_{i k m} {p}_{M N\, j l n p} 
\\
&{}
- \mc{V}_{mn}{}^M {\cal V}^{k l\,N} {\Omega}_{k p} {\e}^{q} {\chi}_{l q r} {p}_{M N}\,^{m n p r} 
- \frac{1}{12}\, {\cal M}^{M N} {\Omega}_{i j} {\e}^{i} {\chi}_{k l m} {p}_{M N}\,^{j k l m} 
\\
&{}
- \frac{1}{2}\, \mc{V}_{mn}{}^M {\cal V}^{k l\,N} {\Omega}_{k p} {\e}^{p} {\chi}_{l q r} {p}_{M N}\,^{m n q r} 
+  \mc{V}_{mn}{}^M {\cal V}^{k l\,N} {\Omega}_{k p} {\e}^{m} {\chi}_{l q r} {p}_{M N}\,^{n p q r}
\;.
\end{aligned}\end{equation}
Finally, one employs the relations
\begin{equation}\begin{aligned}
\mc{V}_{mn}{}^M {\cal V}^{k l\,N} {\Omega}_{k p} {\e}^{q} {\chi}_{l q r} {p}_{M N}\,^{m n p r} 
=&
-\frac13\,
{\cal V}^{i\llbracket j\,M}{\cal V}^{l\rrbracket k\,N} \Omega_{ik}\Omega_{lq} \e^q \Omega^{rs}
\chi_{rmn} p_{MN}{}^{mn}{}_{js}
\\
&{}
- {\cal V}^{i\llbracket j\,M}{\cal V}^{l\rrbracket k\,N} \Omega_{ik} \e^r
\chi_{lmn} p_{MN}{}^{mn}{}_{jr}
\;,
\end{aligned}\end{equation}
and
\begin{equation}\begin{aligned}
{}
\mc{V}_{mn}{}^M {\cal V}^{ k l \,N} {\Omega}_{p q} {\e}^{p} {\chi}_{k l r} {p}_{M N}\,^{m n q r} 
- 2\,{\cal V}^{ij\,M} {\cal V}^{kl\,N} {\Omega}^{m n} {\e}^{p} {\chi}_{i k m} {p}_{M N\, j l n p} 
- \frac{1}{12}\, {\cal M}^{M N} {\Omega}_{i j} {\e}^{i} {\chi}_{k l m} {p}_{M N}\,^{j k l m} 
\\
=
2\, {\cal V}^{i\llbracket j\,M}{\cal V}^{l\rrbracket k\,N} \Omega_{ik} \e^r
\chi_{lmn} p_{MN}{}^{mn}{}_{jr},
\nonumber
\end{aligned}\end{equation}
of which both r.h.s. vanish due to the section constraint.
Together, we conclude that the expression (\ref{rescomm}) contains
\begin{equation}\begin{aligned}
&\longrightarrow&
\mc{V}_{mn}{}^M {\cal V}^{kl\,N} {\Omega}_{p [q} {\e}^{p} {\chi}_{k l r]} {p}_{M N}\,^{m n q r} 
- \frac{1}{12}\, {\cal M}^{M N} {\Omega}_{i j} {\e}^{i} {\chi}_{k l m} {p}_{M N}\,^{j k l m} 
\;,
\end{aligned}\end{equation}
which is precisely the contribution from $-\frac14\,{\cal R}^{ijkl}$ from (\ref{Rijkl}).

\end{appendix}


\providecommand{\href}[2]{#2}\begingroup\raggedright\endgroup

\end{document}